\documentclass[11pt]{article}

\usepackage[final]{acl}

\usepackage{times}
\usepackage{latexsym}
\usepackage{todonotes}
\usepackage{amsfonts}
\usepackage{booktabs}
\usepackage{multirow}
\usepackage{caption}
\usepackage{graphicx}
\usepackage{subcaption}
\usepackage{siunitx}
\usepackage{fvextra}
\usepackage{tipa}
\DefineVerbatimEnvironment{PromptBlock}{Verbatim}{
  breaklines=true,
  breakanywhere=true,
  breaksymbolleft={},
  breaksymbolright={},
  fontsize=\small,
  frame=none
}

\usepackage{pifont} % http://ctan.org/pkg/pifont
\newcommand{\cmark}{\ding{51}}%
\newcommand{\xmark}{\ding{55}}%

\usepackage[T1]{fontenc}
\usepackage[utf8]{inputenc}

\usepackage{microtype}

\usepackage{inconsolata}

\usepackage{graphicx}

\title{ChildVox: A Speech, Audio, and Large Audio-Language Model Benchmark in Understanding and Characterizing Sound across Childhood}

\author{
Tiantian Feng\textsuperscript{1} \quad
  Anfeng Xu\textsuperscript{1} \quad
  Xuan Shi\textsuperscript{1} \quad
  Aditya Kommineni\textsuperscript{1} \quad
  Shakhrul Iman Siam\textsuperscript{3} \\
  \textbf{Megan Micheletti\textsuperscript{4}} \quad
  \textbf{Zhonghao Shi\textsuperscript{5}} \quad
  \textbf{Helen Tager-Flusberg\textsuperscript{6}} \quad
  \textbf{Mi Zhang\textsuperscript{3}} \\
  \textbf{Lynn K. Perry\textsuperscript{2}} \quad
  \textbf{Catherine Lord\textsuperscript{4}} \quad
  \textbf{Daniel Messinger\textsuperscript{2}} \quad
  \textbf{Shrikanth Narayanan\textsuperscript{1}}
 \\
 \\
\textsuperscript{1}University of Southern California \quad
\textsuperscript{2}University of Miami \quad
\textsuperscript{3}The Ohio State University \\
\textsuperscript{4}University of California, Los Angeles \quad
\textsuperscript{5}Harvard University \quad
\textsuperscript{6}Boston University \\
\textbf{Correspondence:} \href{tiantiaf@usc.edu}{tiantiaf@usc.edu} % ORCID=0000-0002-2053-9068
}

\begin{document}
\maketitle
\begin{abstract}
We present \texttt{ChildVox}, a novel benchmark for characterizing the diverse acoustic signals through which children communicate. Specifically, \texttt{ChildVox} follows the full developmental trajectory from birth through school age, covering physiological sounds, non-linguistic vocalizations, canonical syllables, and spoken language. \texttt{ChildVox} integrates more than 20 sub-tasks across 17 child-centered audio and speech datasets, enabling systematic cross-corpus and cross-domain comparison. We evaluate a representative range of audio and speech foundation models, including self-supervised, ASR-oriented, and large audio-language models, on tasks including physiological sound classification, vocalization and canonical syllables modeling, and speech quality assessment and recognition. Benchmark results show that \texttt{ChildVox} provides a suite of high-performance models in recognizing a wide range of acoustic signals from children, supporting downstream applications such as characterizing children's language levels and tracking speech production with age. Our benchmark models on public datasets are released at: \url{https://github.com/tiantiaf0627/childvox-release}.
\end{abstract}

\section{Introduction}
\label{sec:intro}

Deep learning and Artificial Intelligence for child-centered speech or voice processing and modeling have long focused on automatic speech recognition (ASR)~\cite{fan2024benchmarking, ying2025benchmarking, singh2026causal}, a general-purpose task for populations with established spoken language skills. 
However, children ``express'' themselves through a wide range of verbal and non-verbal skills at different stages of childhood~\cite{li2026automated}, shaped by their physical~\cite{namasivayam2020speech} and neuro-developmental status~\cite{lord2018autism} and educational backgrounds~\cite{lesaux2003development}. 
Therefore, tasks like ASR cover only a subset of children's communicative skills and life experiences, and do not fully capture signals critical for understanding development, language level, health, and behaviors across childhood. 

This limitation is amplified for children who have not yet fully acquired socially recognized spoken language, or who experience speech sound disorders (SSD) and other conditions that impact speech production. In early childhood, for example, communication is often dominated by non-spoken forms, including vocalized events (e.g., cries). Moreover, physiological sounds (e.g., cardiac sounds~\cite{oliveira2021circor}) can carry critical information when verbal or non-verbal cues are insufficient for understanding a child's condition.

In this work, we therefore rethink what constitutes speech, voice, or language in the context of child development. We introduce \texttt{ChildVox}, an audio benchmark that extends conventional ASR modeling to broader forms of children's voices'', enabling the characterization of sounds across childhood experiences for child-centered applications. 
We broadly view ``voice'' in children as embodied communication that includes not only spoken language, but also vocalization, canonical syllables, and physiological sounds, all of which work together to convey meaning and express internal states. 
This definition calibrates communication skills to the developmental stage, recognizing that how children interact with the world evolves as their linguistic, motor, and cognitive skills develop.

\begin{figure*} {
    \centering
    
    \includegraphics[width=0.96\linewidth]{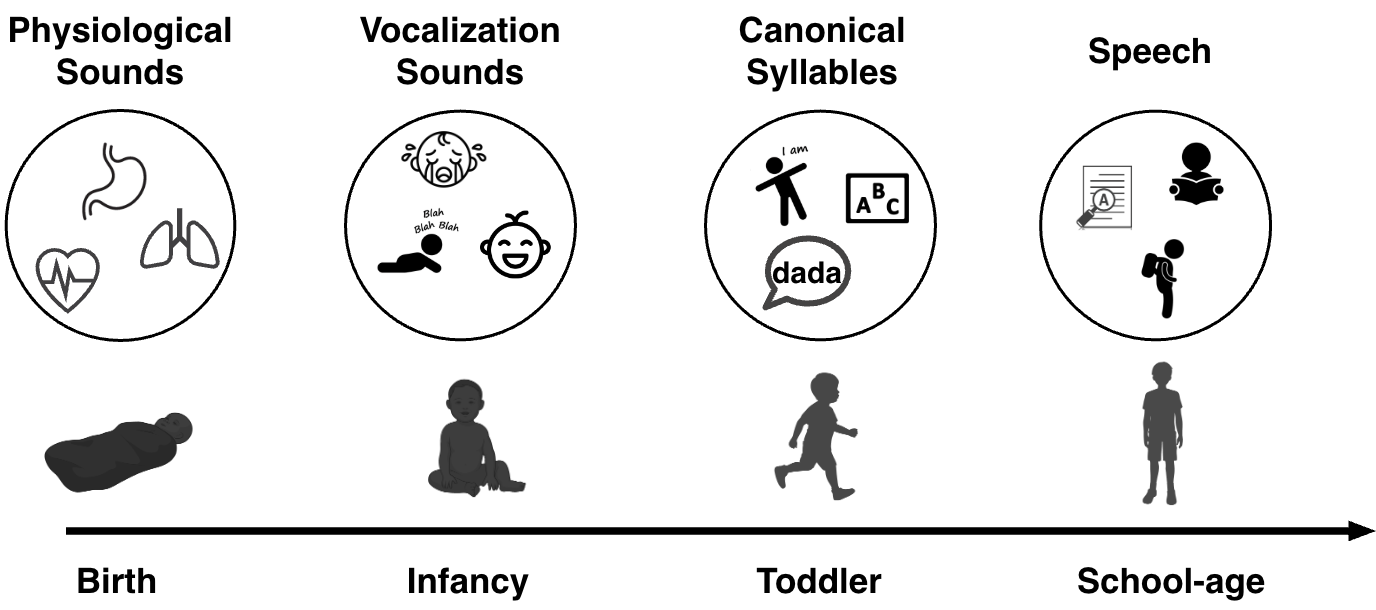}
    \vspace{-2mm}
    \caption{Overview of the wide range of ways that children ``express'' themselves. \texttt{ChildVox} benchmarks audio, speech, and large audio-language models on physiological sounds, vocalizations, canonical syllables, and speech.}
    \vspace{-2mm}
    \label{fig:ChildVox}
} \end{figure*}

Our proposed \texttt{ChildVox} benchmark offers unique contributions compared to prior works: (1) Unlike previous studies that primarily focus on children's ASR, \texttt{ChildVox} considers the full developmental trajectory from birth through school age, covering physiological sounds, vocalizations, canonical syllables, and speech within a single unified benchmark. Here, we specifically refer to ``unified'' as the integration and evaluation of heterogeneous child-centered acoustic signals drawn from different datasets and developmental stages. (2) \texttt{ChildVox} integrates more than 20 sub-tasks across 17 child-centered audio and speech datasets into a consistent evaluation protocol. (3) We evaluate a representative range of self-supervised, ASR-oriented, and large audio-language models, showing that different pre-training setups offer complementary measures of child-centered acoustic signals. 
Moreover, models provided in \texttt{ChildVox} can outperform the frontier proprietary models like Gemini 3.5 Flash in zero-shot settings on the five selected tasks.
Extensive experiments show that \texttt{ChildVox} provides a robust platform for supporting downstream applications in understanding children's audio and spoken language across developmental contexts.

\section{Related Works}
\subsection{Child ASR Benchmarks}

Several recent studies have proposed benchmark-style evaluations for children's ASR, where many of these works consistently show substantial degradation when adult-trained ASR systems are applied to children's speech, due to developmental acoustic variability, pronunciation differences, and conversational dynamics.
For example, \citet{fan2024benchmarking, ying2025benchmarking} evaluate supervised and self-supervised speech foundation models for child ASR, analyzing the effects of model scaling, training paradigms, and datasets. 
Other studies examine ASR robustness of speech from children with developmental delays, highlighting challenges from atypical vocal behaviors~\cite{mcgonigle2024benchmarking}. 
Recently, causal analyses of children's ASR errors further quantify the contributions of physiological, cognitive, and environmental factors to recognition failures~\cite{singh2026causal}.

\begin{table*}[ht]
    \centering

    \caption{Summary of the audio and speech data used in the \texttt{ChildVox} benchmark. The proposed benchmark includes more than 20 sub-tasks related to physiological sound classification, vocalization classification, canonical syllables classification, and speech recognition and classification.}
    \vspace{-1mm}
    \resizebox{\linewidth}{!}{
    \begin{tabular}{lcccccc}

        \toprule
        \multicolumn{1}{c}{\textbf{Dataset}} & 
        \multicolumn{1}{c}{\textbf{Evaluation Tasks}} & 
        \multicolumn{1}{c}{\textbf{Labels}} & 
        \multicolumn{1}{c}{\textbf{Data License}}
        \\
        \midrule

        \multirow{1}{*}{CirCor~\cite{oliveira2021circor}} & Murmur Detection & Absent, Present, Unknown & Open Database \\
        \cmidrule(lr){2-4}

        \multirow{3}{*}{ICBHI~\cite{rocha2019open}} & Crackle Detection & Crackles, No crackles & \multirow{3}{*}{Open Database} \\
        
         & Wheeze Detection & Wheezes, No wheezes & \\
         
        & Respiratory Condition & Healthy, COPD, Other condition & \\
        \cmidrule(lr){2-4}
        
        \multirow{2}{*}
        {SPRSound~\cite{zhang2022sprsound}} & \multirow{2}{*}{Respiratory Sound} & Normal, CAS, DAS, & 
        \multirow{2}{*}{Not Specified} \\
        
        & & CAS \& DAS, Poor Quality &\\

        \midrule

        Donate-a-cry & Cry Classification & Hunger, Other & Open Database
        \\
        \cmidrule(lr){2-4}
        
        \multirow{1}{*}{What's in a cry?~\cite{lockhart2023infant}} & Cry Classification & Hunger, Loneliness, Discomfort & Not Specified \\
        \cmidrule(lr){2-4}
        
        \multirow{4}{*}{AudioSet~\cite{gemmeke2017audio}} & & Child speech, Babbling, Giggle, & \multirow{4}{*}{CC-BY-4.0} \\
        & Child Sound & Children shouting, Baby laughter, & \\
        & Classification & Baby cry, Whimper, Child singing, & \\
         & & Children playing, Music for children & \\
        \midrule

        \multirow{1}{*}{ReCANVo~\cite{johnson2023recanvo}} & \multirow{2}{*}{Affective Status} & Delighted, Dysregulated, Frustrated, & \multirow{2}{*}{Not Specified} \\
        (include adult audio data) & & Request, Selftalk, Social & \\
        \cmidrule(lr){2-4}
        
        \multirow{2}{*}{BabbleCor~\cite{cychosz2021vocal}} & Vocal Development & Crying, Laughing, Canonical, & \multirow{2}{*}{Customized License} \\
        & Classification & Non-Canonical, Junk & \\
        \cmidrule(lr){2-4}
        
        SpeechMaturity & Vocal Development & Crying, Laughing, Canonical, & \multirow{2}{*}{Customized License} \\
        
        ~\cite{hitczenko2025speech} & Classification & Non-Canonical, Junk & \\
        
        \midrule
        
        \multirow{1}{*}{C-BESD~\cite{rao2024children}} & Emotion Recognition & Anger, Happy, Neutral, Sad & Not Specified \\
        \cmidrule(lr){2-4}
        
        \multirow{2}{*}{PERCEPT-R~\cite{benway2022percept}} & \multirow{2}{*}{Pronunciation} & \multirow{2}{*}{Rhotic, Derhotic} & PhonBank License \\

        & & & (CC-BY-NC-SA 3.0) \\

        \cmidrule(lr){2-4}

        \multirow{3}{*}{\shortstack[l]{SpeechOcean762~\cite{zhang2021speechocean762}\\ (includes adult audio data)}} & Prosody & Poor, Nearly correct, Correct intonation & \multirow{3}{*}{CC-BY-4.0} \\
         & Fluency & Little fluent, Fluent in general, Fluent & \\
         & Pronunciation & Poor or Understandable, Good, Excellent & \\
        \cmidrule(lr){2-4}

        \multirow{2}{*}{UltraSuite~\cite{eshky2018ultrasuite}} & \multirow{2}{*}{Speech Articulator} & Bilabial, Labiodental, Dental, Glottal & \multirow{2}{*}{CC-BY-NC-4.0}
        \\
        
        & & Labiovelar, Alveolar, Palatal, Velar &
        \\
        \cmidrule(lr){2-4}

        Natural Language Sampling (NLS) & Diarization & Speaker labels and timestamps & \multirow{2}{*}{Private} \\
        
        ~\cite{butler2022remote} & Speech Intelligibility & Intelligible, Unintelligible, Vocalization & \\
        \cmidrule(lr){2-4}

        \multirow{2}{*}{In-house ADOS2-Mod3} & Diarization & Speaker labels and timestamps & \multirow{2}{*}{Private} \\
        
        & ASR & Speech Transcript & \\
        \cmidrule(lr){2-4}
        
        MyST~\cite{pradhan2024my} & ASR & Speech Transcript & Customized License
        \\
        \cmidrule(lr){2-4}
        TinyVox~\cite{lavechin2026babar} & ASR & Phoneme Transcript & Not Specified
        \\

        \bottomrule

    \end{tabular}
    }
    \label{tab:datasets}
    \vspace{-1mm}
\end{table*}

\subsection{Child Speech Benchmarks beyond ASR}

Apart from ASR, recent work has applied speech foundation models to a range of child-centered tasks, showing that self-supervised representations capture meaningful developmental information from children's speech and infant vocalizations~\cite{li2024analysis}.
Other studies have explored speech foundation models for child-adult speaker diarization in naturalistic dyadic interactions~\cite{xu2024exploring}, spoken language development analysis in children with ASD~\cite{xu2023understanding}, and phoneme recognition for language learning applications~\cite{medin2025self}. Additionally, layer-wise analyses from a recent work further reveal that different model layers encode distinct developmental information~\cite{sinha2026study}. Collectively, these studies show the growing potential of speech foundation models beyond ASR in children's speech.

\section{ChildVox Benchmark}
\label{sec:benchmark}
%This section introduces the \texttt{ChildVox} benchmark, covering its evaluation tasks, datasets, and label definitions. 
Figure~\ref{fig:ChildVox} presents the overview of benchmark design to characterize the range of ways that children ``express'' themselves from birth through school age.

\subsection{Benchmark Overview}

Unlike conventional ASR-centric benchmarks on children's speech, \texttt{ChildVox} captures the diverse spectrum of child-produced and child-related acoustic signals that emerge throughout development: physiological sounds, such as cardiac sounds that provide markers of health conditions; vocalizations, such as crying in infancy; canonical syllables and emerging language behaviors in toddlers, such as proto-speech that reflect the progressive development of communicative skills; and the more complex social communicative behaviors of school-age children associated with spoken language interaction in everyday interactions. Collectively, this developmental framing across childhood positions the \texttt{ChildVox} as an ecologically grounded platform for advancing child-centered audio intelligence.

\subsection{Benchmark Tasks and Datasets}

\texttt{ChildVox} evaluates four broad categories of child-centered audio datasets: physiological sounds, vocalizations, canonical syllables, and speech. The information regarding evaluation tasks, labels, and data license is provided in Table~\ref{tab:datasets}. The details about each dataset are described in the Appendix~\ref{app:dataset}.

% on a comprehensive collection of child-centered audio datasets covering four broad categories: 

\paragraph{Physiological Sound Classification}
\texttt{ChildVox} includes three benchmark datasets for pediatric physiological sound classification covering both cardiac and respiratory assessment. 
The benchmark targets four tasks: murmur detection, crackle detection, wheeze detection, and respiratory condition classification. Murmur detection is evaluated on CirCor~\cite{oliveira2021circor}, a collection of 5,272 pediatric cardiac auscultation recordings, while crackle detection, wheeze detection, and respiratory condition classification are evaluated on ICBHI~\cite{rocha2019open} and SPRSound~\cite{zhang2022sprsound}, two respiratory sound datasets consisting of pediatric and general patient populations.

\paragraph{Vocalization Event Classification}
\texttt{ChildVox} targets two vocalization-related tasks: generic child sound and infant cry classification. Child sound classification covers ten child-related audios: speech, babble, giggle, shout, laughter, cry, whimper, sing, play, and music for children. This is drawn from AudioSet~\cite{gemmeke2017audio}, a large-scale ontology-driven collection of YouTube data. Cry cause classification is evaluated on two infant cry corpora: Donate-a-Cry~\footnote{\url{https://github.com/gveres/donateacry-corpus}}, where we classify hunger versus other, given its imbalanced distribution, and What's in a cry?~\cite{lockhart2023infant} contains labels: hunger, loneliness, and discomfort.

\paragraph{Canonical Syllables Classification}

We experimented with ReCANVo~\cite{johnson2023recanvo}, a database of affective nonverbal sounds recorded in everyday interactions from minimally speaking children and adults with neurodevelopmental conditions. This corpus supports affective status classification (e.g., dysregulated). 
Although ReCANVo includes recordings from both children and adults, it remains a unique resource for studying affective nonverbal vocalizations in children.
BabbleCor~\cite{cychosz2021vocal} and SpeechMaturity~\cite{hitczenko2025speech,zhang25r_interspeech} target early vocal development in children, supporting vocal development classification across cry, laugh, canonical, non-canonical, and junk.

% \subsubsection{Speech Classification and Recognition}

\paragraph{Speech Quality Assessment}
We include three datasets evaluating speech production, pronunciation, and articulation. PERCEPT-R~\cite{benway2022percept} (with PhonBank~\cite{rose2014phonbank}) contains recordings from children with typical speech development and speech sound disorders, emphasizing rhotic production in American English. SpeechOcean762~\cite{zhang2021speechocean762} provides 5,000 English utterances from 250 non-native speakers (half children), each annotated by five expert raters for pronunciation accuracy, fluency, and prosody.
Similar to the ReCANVo dataset, although SpeechOcean762 includes half adult speech data, it remains one of the largest and well-annotated corpora to evaluate children's speech quality.
UltraSuite~\cite{eshky2018ultrasuite} pairs synchronized ultrasound tongue imaging with acoustic recordings for research on child speech disorders. Inspired by \citet{ribeiro2021exploiting}, we use it for articulator classification across eight places of articulation (e.g., velar) on word-level audios.

\begin{figure}[t] {
    \centering
    
    \includegraphics[width=\linewidth]{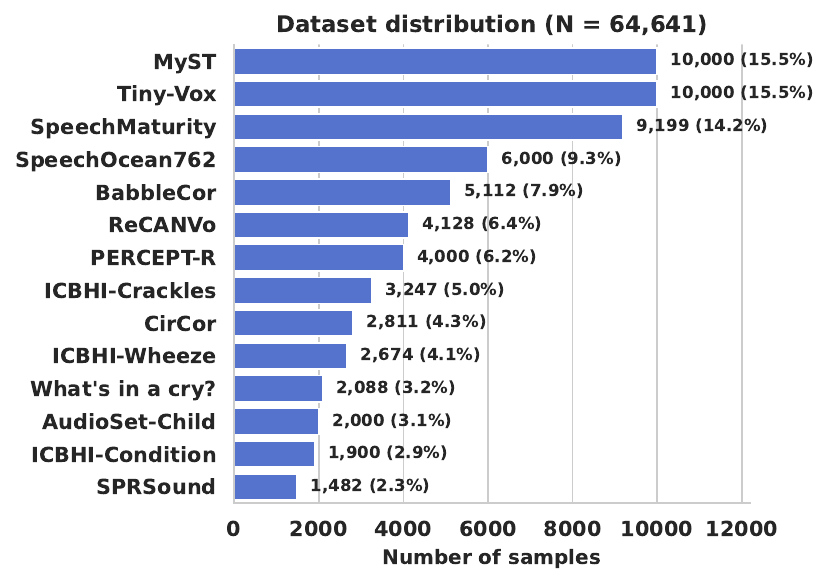}
    \vspace{-7mm}
    \caption{Training data distribution of each sub-dataset in \texttt{ChildVox-Balanced} dataset for training LALMs.}
    \vspace{-3.5mm}
    \label{fig:ChildVox_balance}
} \end{figure}

\begin{table*}[t]
    \begin{minipage}[b]{0.56\linewidth}
        \resizebox{\linewidth}{!}{
            \begin{tabular}{lcc}
            
            \multirow{2}{*}{\textbf{Pre-trained Model}} & 
            \multirow{2}{*}{\textbf{Pre-Training Objectives}} & 
            \textbf{Child-centered}  \\ 

            & & \textbf{Audio Pre-Training}  \\ 
    
            \midrule
            \textbf{SSAST-Base} & \multirow{5}{*}{Self-supervised} & \xmark \\
            \textbf{voc2vec-HuBERT} & & \xmark \\
            \textbf{wav2vec-LB} & & \cmark \\
            \textbf{BabyHuBERT} & & \cmark \\
            \textbf{WavLM Large} & & \xmark \\
            \midrule
            \textbf{Whisper Base} & \multirow{3}{*}{VAD, ASR, LID} & \xmark \\
            \textbf{Whisper Small} & & \xmark \\
            \textbf{Whisper Large V3} & & \xmark \\
            \midrule
            \textbf{Qwen2-Audio Instruct} & Multi-stage Uni-modal & \xmark \\ 
            \textbf{AudioFlamingo 3} & and Multi-modal Learning & \xmark \\
            \bottomrule
            \end{tabular}
        }
        \vspace{0mm}
        \caption{Summary of audio and speech foundation models used in the ChildVox benchmark. We mark whether the pre-trained models are pre-trained with child-centered audio.}
        \label{tab:foundation_models}
    \end{minipage} %
    \hfill
    \begin{minipage}[b]{0.4\linewidth}
        \includegraphics[width=\linewidth]{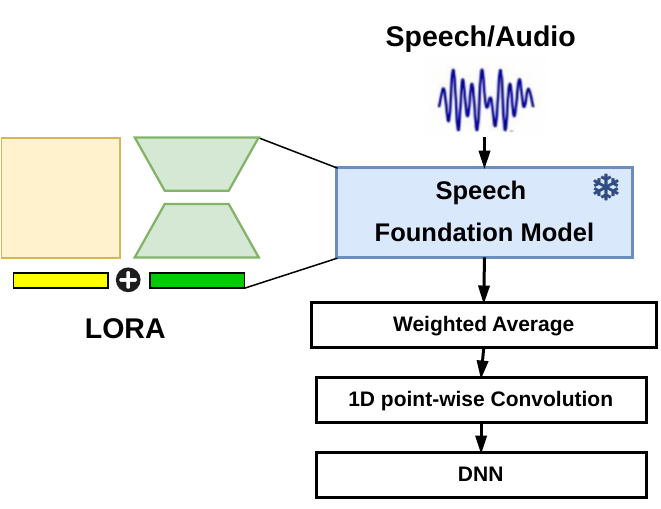}
        \vspace{-6.5mm}
        \captionof{figure}{LoRA fine-tune in most encoder-based pre-trained models except SSAST, BabyHuBERT and wav2vec-LB.}
        \label{fig:lora}
    \end{minipage}
    \vspace{-3mm}
\end{table*}

\paragraph{Adult-child Speaker Diarization} 

A core component of analyzing child speech in interaction is speaker diarization, which detects speech segments and labels ``who spoke when''. Here, we include two in-house datasets to benchmark the speaker diarization. Specifically, the Natural Language Sampling (NLS)~\cite{butler2022remote} dataset includes child-parent conversations collected from remote dyadic interactions involving minimally verbal children with ASD. The corpus includes 73 English-speaking child-parent sessions with detailed child/ adult speaker and speech intelligibility labels. Moreover, the ADOS2-Mod3 dataset~\cite{xu2026end} consists of recordings from more than 100 children participating in Autism Diagnostic Observation Schedule-Module3~\cite{lord2008autism} assessments. Within this dataset, interactions related to emotion and social-difficulty segments are annotated with speaker labels, which are kept for the fine-tuning experiment.

\paragraph{Automatic Speech Recognition} 
We target both word-level and phoneme-level transcription of children's speech. Word-level recognition is evaluated on MyST~\cite{pradhan2024my}, a corpus of elementary school students interacting with a virtual science tutor, and on ADOS2-Mod3, whose transcripts are annotated with emotion and social-difficulty segments. Phoneme-level recognition is evaluated on TinyVox~\cite{lavechin2026babar}, which provides over half a million phoneme-level transcriptions across multiple languages. Here, we focus on the English subset with one item.

\subsection{\texttt{ChildVox}-Balanced Dataset}

Given the substantial aggregate scale of the combined datasets, we construct \texttt{ChildVox}-Balanced, a curated subset with labels balanced within each classification task, to support practical and reproducible supervised fine-tuning of Large Audio Language Models (LALMs) mentioned later. It draws only on publicly accessible resources and limits training and testing samples per label to 2,000 and 50, respectively. For ASR evaluation, we include 10,000/500 utterances from MyST and TinyVox for training/testing. 
Figure~\ref{fig:ChildVox_balance} shows the resulting per-dataset distribution, yielding a total of 64,641 audio and speech recordings across 14 subtasks.

\section{Benchmark Models}

In \texttt{ChildVox}, we evaluate several representative audio, speech, and large audio-language models for benchmark experiments. The details of the experimented pre-trained models are listed in Table~\ref{tab:foundation_models}.

\subsection{Encoder-based Models}

\paragraph{Self-supervised Models} We evaluate five representative self-supervised audio and speech foundation models: \texttt{SSAST}~\cite{gong2022ssast}, \texttt{voc2vec}~\cite{koudounas2025voc2vec}, \texttt{WavLM}~\cite{chen2022wavlm}, \texttt{wav2vec-LB}~\cite{li2023towards}, and BabyHuBERT~\cite{charlot2025babyhubert}. \texttt{SSAST}, \texttt{voc2vec}, and \texttt{WavLM} apply self-supervised learning to learn representations from generic audio events, non-verbal vocalizations, and adult-centered speech data, respectively. These pre-trained models have shown strong performance on audio event detection, speech recognition, and speaker diarization tasks. Moreover, BabyHuBERT and wav2vec-LB are two pretrained speech models trained primarily on child-centered audio and speech data using HuBERT and wav2vec pre-training, respectively.

\paragraph{ASR Models} 

We primarily evaluate models from the Whisper family, an encoder-decoder architecture trained on large-scale multilingual speech primarily for ASR. Apart from ASR, Whisper encoders have shown strong transferability across speech processing problems, such as child-adult speaker diarization~\cite{xu2024exploring}. 
To explore the effect of model scale, we evaluate encoder embeddings from three Whisper variants: Whisper-Base, Whisper-Small, and Whisper-Large-v3. Moreover, we include Parakeet-TDT~\cite{rekesh2023fast, xu2023efficient} for fine-tuned ASR tasks.

\begin{table*}[ht]
    \centering
    \caption{Benchmark performance across physiological sounds, vocalization, canonical syllables, and speech from children on each complete dataset. 
    voc2vec-H, WavLM-L, Whisper-L are voc2vec-HuBERT, WavLM-Large, Whisper-Large, respectively. SO represents the SpeechOcean dataset. Additional results on Whisper-base and Whisper-small are reported in the Appendix~\ref{sec:additional_benchmark}.
    We report the average Macro-F1 score (higher the better), and the best and second-best results per row are in \textbf{bold} and \underline{underlined}, respectively.}
    \small
    \setlength{\tabcolsep}{6pt}
    \begin{tabular}{c l *{6}{S[table-format=1.3]}}
    \toprule
    {Task Category} & {Dataset} & {SSAST} & {voc2vec-H} & {WavLM-L} & {wav2vec-LB} & {BabyHuBERT} & {Whisper-L} \\
    \midrule
     & {CirCor} & \underline{0.636} & 0.563 & \textbf{0.643} & 0.575 & 0.632 & 0.552 \\
     & {ICBHI-crackles}  & \underline{0.644} & {0.626} & 0.612 & 0.625 & \textbf{0.654} & 0.586 \\
     Physiological & {ICBHI-wheezes} & \textbf{0.638} & 0.592 & 0.614 & 0.613 & \underline{0.636} & 0.609 \\
     Sounds & {ICBHI-Condition} & 0.466 & 0.445 & 0.392 & \underline{0.472} & 0.401 & \textbf{0.508} \\
     & {SPRSound} & \textbf{0.448} & 0.368 & \underline{0.439} & 0.189 & 0.359 & {0.414} \\
     % & {SPRSound-ternary} & \underline{0.667} & 0.629 & \textbf{0.702} & 0.581 & 0.575 & 0.666 \\
    \midrule
    \multirow{3}{*}{Vocalization}
     & {AudioSet-Child} & \textbf{0.657} & 0.560 & 0.539 & 0.606 & \underline{0.632} & 0.619 \\
     & {What's in a cry?} & 0.300 & 0.303 & {0.321} & {0.308} & \textbf{0.326} & \underline{0.323} \\
     & {donate-a-cry}  & 0.473 & \underline{0.486} & 0.468 & 0.424 & 0.430 & \underline{0.502} \\
    \midrule
    \multirow{3}{*}{\shortstack{Canonical\\Syllables}}
     & {ReCANVo} & \underline{0.444} & 0.416 & {0.423} & 0.425 & \textbf{0.446} & 0.359 \\
     & {Speech Maturity}  & {0.686} & 0.640 & 0.620 & \underline{0.696} & \textbf{0.722} & {0.640} \\
     & {BabbleCor} & 0.569 & 0.514 & 0.485 & \underline{0.605} & \textbf{0.667} & {0.594} \\
    \midrule
    \multirow{7}{*}{Speech}
     & {NLS-Intelligibility} & 0.588  & 0.548 & 0.603 & 0.602 & \underline{0.614} & \textbf{0.661} \\
     & {Percept-R}  & 0.789 & 0.788 & \underline{0.813} & 0.805 & \textbf{0.823} & 0.811 \\
     & {Ultrasuite} & 0.850  & 0.635 & \underline{0.882} & 0.768 & 0.818 & \textbf{0.888} \\
     & {SO-Fluency}  & 0.601 & 0.554 & \underline{0.613} & 0.516 & \textbf{0.627} & \textbf{0.627} \\
     & {SO-Accuracy} & 0.564  & 0.557 & \underline{0.625} & 0.575 & {0.598} & \textbf{0.649} \\
     & {SO-Prosody}  & 0.613 & 0.613 & \underline{0.705} & 0.585 & 0.669 & \textbf{0.715} \\
     & {C-BESD}  & \underline{0.753} & 0.566 & \textbf{0.892} & 0.658 & 0.665 & 0.532 \\
    \bottomrule
    \end{tabular}
    \label{tab:benchmark_results}
    \vspace{-3mm}
\end{table*}

\subsection{Large Audio-Language Model}

We include two widely used Large Audio-Language Models (LALMs) in our benchmark: Qwen2-Audio-Instruct~\cite{chu2024qwen2} and Audio Flamingo 3 (AF3)~\cite{ghosh2026audio}. Both models unify diverse audio understanding across speech, environmental sounds, and music. Specifically, Qwen2-Audio combines a Whisper-large-v3 audio encoder with the Qwen-7B language model backbone. 
Similar to Qwen2-Audio, the audio encoder of AF3 is initialized from Whisper-Large-v3.

\section{Experiments}

\subsection{Data Preprocessing}

For all datasets included in \texttt{ChildVox}, audio samples are resampled to 16 kHz to match the sampling rate of all pre-trained models. Across most experiments, we restrict a minimum input duration of 200 ms to maintain sufficient acoustic information for downstream evaluation. For most datasets without predefined training, development, and test partitions, we adopt a 5-fold cross-validation setup in the experiment. More details, such as the prompt design for the SFT experiments with LALM and data filtering, are in the Appendix~\ref{app:dataset} and ~\ref{sec:prompt}.

\subsection{Modeling Details}

The details of training settings, like learning rate and training epochs, are listed in the Appendix~\ref{sec:training}.

\paragraph{Model Fine-tuning} 

Because the experimental pretrained models differ in architecture and implementation, we adopt model-specific fine-tuning strategies. For SSAST, we choose to fine-tune the model using full-model optimization, whereas the pretrained backbone weights of all other models remain frozen. For wav2vec-LB, we use representations from the final hidden layer for downstream fine-tuning. For BabyHuBERT, we learn a weighted combination of representations from all hidden layers for fine-tuning. For voc2vec-HuBERT, WavLM-Large, and the Whisper-family models, we adopt parameter-efficient fine-tuning through LoRA. Specifically, we apply LoRA with a rank of 64 to the feedforward layers for voc2vec, WavLM, and Whisper as shown in Figure~\ref{fig:lora}. For LALMs, LoRA modules with rank 64 are inserted into the query, key, value, down-projection, and up-projection layers of the language model, and the feedforward layers of the audio encoder.

\paragraph{Evaluation}
We report utterance-level Macro-F1 on the test set for most classification tasks in the benchmark. For speaker diarization, we report diarization error metrics, while ASR and phoneme recognition are evaluated using word error rate (WER) and phoneme error rate (PER), respectively. Results for encoder-based models are reported as averages across multi-folds in Appendix~\ref{app:dataset}. Specifically, we use the default test set from datasets like SpeechOcean762 and MyST.

\section{Benchmark Results}

\subsection{Encoder-based Model Results}

Table~\ref{tab:benchmark_results} presents the results of the encoder-based model on the complete \texttt{ChildVox} benchmark data. 

\paragraph{Physiological Sound Classification}
Performance on physiological sound classification varies across models and datasets. For example, WavLM-Large achieves the best result on CirCor with a Macro F1 score of 0.643, while SSAST performs best on ICBHI-wheezes and SPRSound. BabyHuBERT achieves the highest score on ICBHI-crackles with a Macro F1 score of 0.654, whereas Whisper-Large performs best on ICBHI-Condition. 
Overall, models pretrained on general audio or self-supervised speech objectives perform strongly across most tasks in physiological sound classification. 
These results suggest that models learned with generic audio representation may better capture fine-grained spectral and temporal characteristics present in physiological signals. 
In contrast, Whisper-based models show comparatively lower performance on most physiological sound tasks except on ICBHI-Condition, indicating that speech-oriented pretraining may not provide benefits for physiological sound classification, particularly in children.

\begin{table}[t]
    \centering
    \caption{Diarization and ASR performance. We report Diarization Error Rate (DER) and Word Error Rate (WER), where the lower is better. Best results per column in \textbf{bold}, second-best \underline{underlined}.}
    \label{tab:diar_asr_results}
    \small
    \setlength{\tabcolsep}{6pt}
    \begin{tabular}{l *{4}{S[table-format=2.2]}}
    \toprule
    & \multicolumn{2}{c}{DER ($\downarrow$)} & \multicolumn{2}{c}{WER ($\downarrow$)} \\
    \cmidrule(lr){2-3} \cmidrule(lr){4-5}
    {Model} & {NLS} & {ADOS} & {MyST} & {ADOS} \\
    \midrule
    WavLM-Large    & 22.10 & 45.60 & 16.56 & 59.34 \\
    Whisper-Base   & 24.20 & 48.20 & 16.99 & 55.85 \\
    Whisper-Small  & \underline{18.70} & \underline{44.00} & 15.93 & 59.91 \\
    Whisper-Large  & \textbf{17.70} & \textbf{42.50} & \textbf{14.80} & \textbf{40.20} \\
    Parakeet-TDT   & {--}  & {--}  & \underline{15.82} & \underline{45.60} \\
    \bottomrule
    \end{tabular}
    \vspace{-2mm}
\end{table}

\paragraph{Vocalization and Canonical Syllables}
Similar to the physiological sound classification, for vocalization and canonical-syllable classification, no single model consistently dominates across datasets. SSAST achieves the best performance on AudioSet-Child and competitive performance in ReCANVo and Speech Maturity, while BabyHuBERT performs best on What's in a cry?, ReCANVo, Speech Maturity, and BabbleCor. Whisper-Large achieves the highest score on donate-a-cry and remains competitive on several other tasks. These results suggest that general audio pretraining is particularly effective for broad child-related acoustic events, whereas pre-trained representations learned from child-relevant speech or sound may better capture early vocal and syllabic patterns. Nevertheless, cry-cause classification remains particularly challenging under the current benchmark setting, with relatively low performance across all evaluated models.

\begin{table*}[t]
    \centering
    \small
    \setlength{\tabcolsep}{3pt}
    \caption{Benchmark performance on \texttt{ChildVox-Balanced} test set using encoder-based models and LALMs. We report PER and WER (lower the better) for TinyVox and MyST, respectively. We report the Macro-F1 score (higher the better) on the remaining sets. SO refers to the SpeechOcean762 dataset.}
    \vspace{-2mm}
    \begin{tabular}{lcccccccccc}
    \toprule
    % \cmidrule(lr){2-4}\cmidrule(lr){5-7}\cmidrule(lr){8-9}\cmidrule(lr){10-14}
    \multirow{2}{*}{\textbf{Model}} & \multirow{1}{*}{CirCor} & \multirow{1}{*}{SPRSound}
    & \multirow{1}{*}{AudioSet} & \multirow{1}{*}{ReCANVo} & Speech & \multirow{1}{*}{PERCEPT-R}
    & SO-Prosody & \multirow{1}{*}{TinyVox} & \multirow{1}{*}{MyST} \\

    & ($\uparrow$) & ($\uparrow$) & ($\uparrow$) & ($\uparrow$) & Maturity ($\uparrow$) & ($\uparrow$) &  ($\uparrow$) & ($\downarrow$) & ($\downarrow$) \\
    
    \midrule
    SSAST & 0.543 & 0.508 & \textbf{0.638} & \textbf{0.500} & {0.735} & 0.800 & 0.644 & - & - \\
    BabyHuBERT & \textbf{0.663} & 0.445 & 0.635 & 0.433 & \textbf{0.786} & 0.809 & 0.745 & - & - \\
    
    WavLM-Large & {0.571} & \textbf{0.518} & 0.538 & 0.448 & 0.612 & \textbf{0.843} & 0.726 & - & 0.147 \\
    Whisper-Base & 0.349 & 0.362 & 0.605 & 0.373 & 0.663 & 0.769 & 0.745 & - & 0.147 \\
    Whisper-Large   & 0.402 & 0.452 & 0.636 & 0.370 & 0.689 & 0.815 & \textbf{0.759} & - & \textbf{0.128} \\
    \midrule
    AudioFlamingo3 & \textbf{0.452} & 0.162 & 0.089 & 0.157 & 0.069 & 0.323 & 0.400 & 0.958 & 0.274 \\
    Qwen2-Audio & 0.397 & \textbf{0.541} & \textbf{0.699} & \textbf{0.514} & \textbf{0.726} & \textbf{0.779} & \textbf{0.671} & \textbf{0.602} & \textbf{0.133} \\
    \bottomrule
    \end{tabular}
    \label{tab:lalm}
    \vspace{-2mm}
\end{table*}

% \paragraph{Speech Classification and Recognition}

\paragraph{Speech Quality Assessment}

For children's speech quality assessment, Whisper-Large achieves the best performance on NLS-Intelligibility, UltraSuite, and all three SpeechOcean762 sub-tasks, including fluency, accuracy, and prosody. On the other hand, BabyHuBERT performs best on Percept-R, while WavLM-Large substantially outperforms the other models on C-BESD for emotion classification. These findings suggest the benefits of large-scale multilingual ASR pretraining for predicting children's speech intelligibility, articulation, and overall speech quality. At the same time, the strong results from BabyHuBERT and WavLM-Large indicate that self-supervised representations pre-trained from child-centered speech recordings remain valuable for tasks involving pronunciation and speech emotion. Overall, these results indicate that speech-based pre-training benefits the classification of children's speech, despite such models being trained predominantly on adult data.

\paragraph{Speaker Diarization and Speech Recognition}
Table~\ref{tab:diar_asr_results} reports diarization and ASR performance across the NLS, ADOS2-Mod3, and MyST datasets. For speaker diarization, Whisper-Large achieves the lowest error rates on both NLS (17.70 DER) and ADOS2-Mod3 (42.50 DER). For ASR, Whisper-Large also obtains the best results, reaching 14.80 WER on MyST and 40.20 WER on ADOS, with Parakeet-TDT ranking second on both ASR benchmarks. Overall, model scale is the dominant factor, with the largest Whisper model leading on the speech recognition and diarization tasks.

\paragraph{Findings} Overall, performance varies substantially across child-centered audio categories, highlighting the diverse acoustic properties of physiological sounds, vocalizations, canonical syllables, and speech. Meanwhile, no single model dominates across all tasks, whereas different encoders capture complementary aspects of these signals.

\subsection{Large Audio Language Models Results}

Table~\ref{tab:lalm} compares the two LALMs against the encoder-based models on the \texttt{ChildVox-Balanced} subset, where two LALMs behave substantially differently in their performance. Qwen2-Audio is consistently competitive with the strongest encoder-based models, achieving the best scores on AudioSet and ReCANVo. It remains close to the best encoder-based models on SpeechMaturity and Percept-R, and obtains a low WER on MyST. In contrast, AudioFlamingo3 performs substantially worse on nearly every task, with particularly low scores on vocalization and canonical syllables classification, such as AudioSet and SpeechMaturity. It also returns an extremely large error rate on TinyVox phoneme recognition (0.958 PER). 

\begin{table}[ht]
    \centering
    \caption{Macro-F1 performance of Gemini 3.5 Flash across sampling temperatures. SO denotes the SpeechOcean dataset. The best results for each dataset are shown in \textbf{bold}.}
    \small
    \setlength{\tabcolsep}{8pt}
    \begin{tabular}{l c c c}
        \toprule
        \multirow{2}{*}{\textbf{Dataset}} &
         & \textbf{Temperature} & \\
         &
        \textbf{0.1} &
        \textbf{0.5} &
        \textbf{1.0} \\
        \cmidrule(lr){1-1} \cmidrule(lr){2-4}
        CirCor      & \textbf{0.449} & {0.364} & 0.342 \\
        SPRSound    & {0.267} & \textbf{0.269} & 0.240 \\
        ReCANVo     & \textbf{0.224} & 0.192 & {0.221} \\
        PERCEPT-R   & \textbf{0.703} & 0.668 & {0.693} \\
        SO-Prosody  & {0.427} & \textbf{0.456} & 0.394 \\
        \bottomrule
    \end{tabular}
    \vspace{-2mm}
    \label{tab:gemini35_temperature}
\end{table}

\paragraph{Error Inspection} Manual inspection indicates that this performance difference is driven by inconsistent instruction-following from the response, where AudioFlamingo3 often returns free-form descriptions (e.g., the audio clip is of a baby laughing'') instead of the requested label. It also tends to summarize or hallucinate on transcription tasks. 
For example, given the reference transcript ``Well, without electricity, an electromagnet would never be possible, because it gives it the magnetism,'' AF3 hallucinates a summary rather than a transcription, producing ``The electromagnet is not working.'' Moreover, we show the confusion matrices of Gemini 3.5 Flash and ChildVox Qwen2-Audio on PERCEPT-R and SpeechOcean in Appendix~\ref {sec:proprietary}.

\subsection{Comparison with Proprietary Baselines}
\begin{figure}[t] {
    \centering
    
    \includegraphics[width=\linewidth]{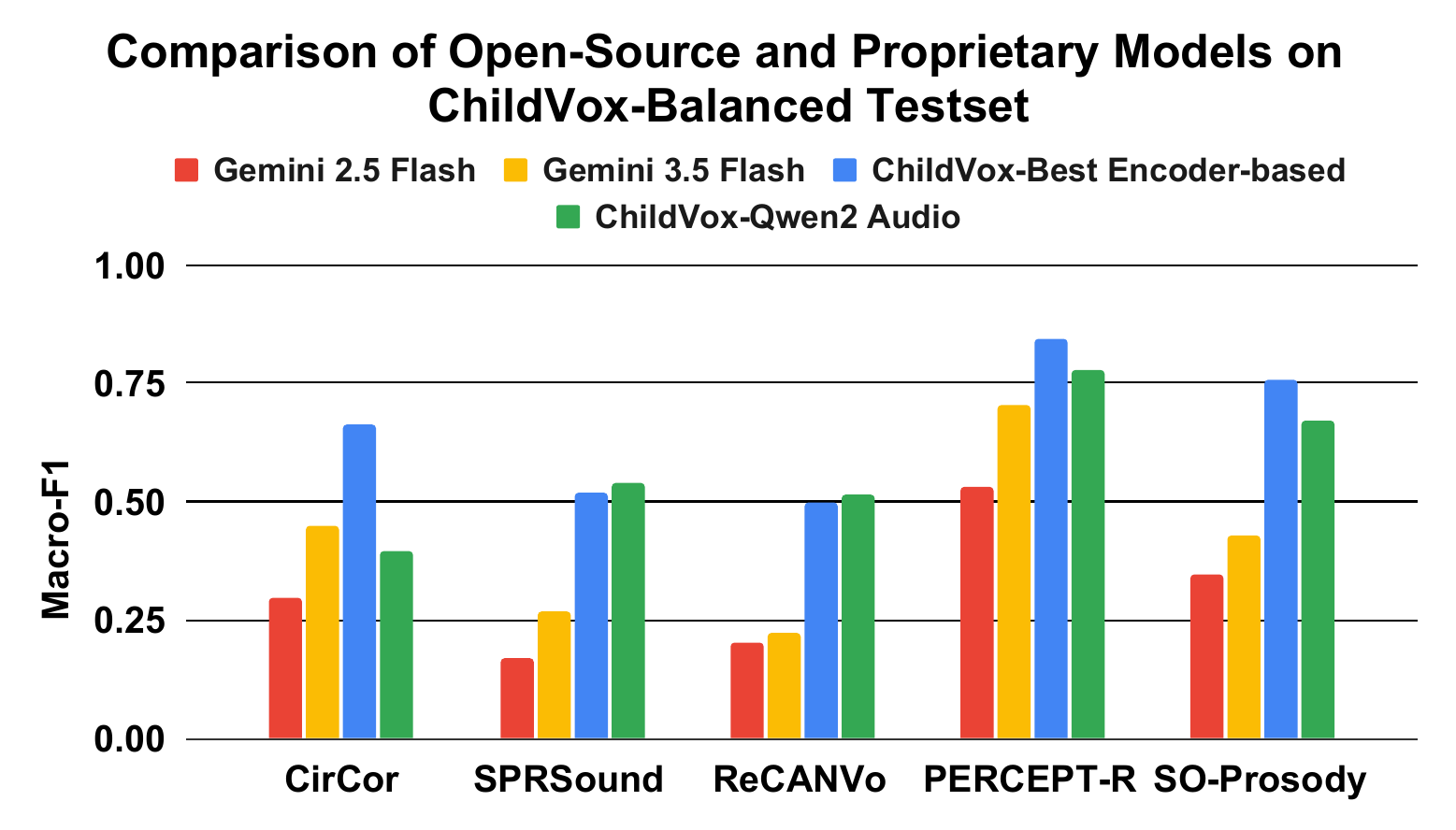}
    \vspace{-6.5mm}

    \caption{Macro-F1 on \texttt{ChildVox-Balanced} test set. Zero-shot proprietary models (Gemini 2.5/3.5 Flash) underperform the \texttt{ChildVox} -trained encoder and Qwen2-Audio baselines across all five datasets.}
    \label{fig:gemini}
    \vspace{-2mm}
} \end{figure}

We compare zero-shot Gemini 2.5 Flash and Gemini 3.5 Flash against our \texttt{ChildVox}-trained encoder (best one) and Qwen2-Audio on five public datasets in Figure~\ref{fig:gemini}. 
We limit this comparison to these five datasets to ensure that no restricted or license-protected data is transmitted to third-party APIs.

\vspace{1mm}
\noindent \textbf{Sensitivity to Temperature} We evaluate both proprietary models across multiple sampling temperatures and compare performance sensitivity to temperature of Gemini 3.5 Flash in Table~\ref{tab:gemini35_temperature}. Specifically, Gemini 3.5 Flash shows sensitivity to the sampling temperature. A temperature of 0.1 provides the strongest overall performance, achieving the best results on CirCor, ReCANVo, and PERCEPT-R, with an average Macro-F1 of 0.414. A temperature of 0.5 performs best on SPRSound and SO-Prosody. At a temperature of 1.0, the average score decreases to 0.378. Overall, lower-temperature decoding is generally more reliable for Gemini 3.5 Flash, although the optimal temperature remains task-dependent. Therefore, we apply a temperature of 0.1 in the remaining analysis.

\vspace{1mm}
\noindent \textbf{Comparing with ChildVox} Overall, specialized models from \texttt{ChildVox} outperform both proprietary baselines in zero-shot settings on five selected task. Specifically, both Gemini models perform particularly poorly on CirCor, SPRSound, and ReCANVo, suggesting that general-purpose LALMs struggle with fine-grained pediatric vocal and physiological audio. In summary, these results highlight the effectiveness of models trained on the \texttt{ChildVox} benchmark. Zero-shot Gemini settings are detailed in the Appendix~\ref{sec:proprietary}.

\section{Exemplar Applications with ChildVox}

Inspired by the previous work~\cite{lavechin2026babar}, we design two experiments using the benchmark models developed from the \texttt{ChildVox} to measure the child's speech production and language development in two different applications. 

\subsection{Characterizing the Language Level}

\paragraph{Application Setup} Our first application is on the in-house NLS dataset described in Section~\ref{sec:benchmark}. In this study, domain experts categorize each child's spoken language level into pre-verbal communication, first words, and word combinations, following the definitions by~\citet{tager2009defining}. Specifically, we refer to these three levels as LL-1, LL-2, and LL-3, respectively. We then apply our speaker diarization estimation models to compare language level differences across three groups.

\begin{figure}[t] {
    \centering
    
    \includegraphics[width=\linewidth]{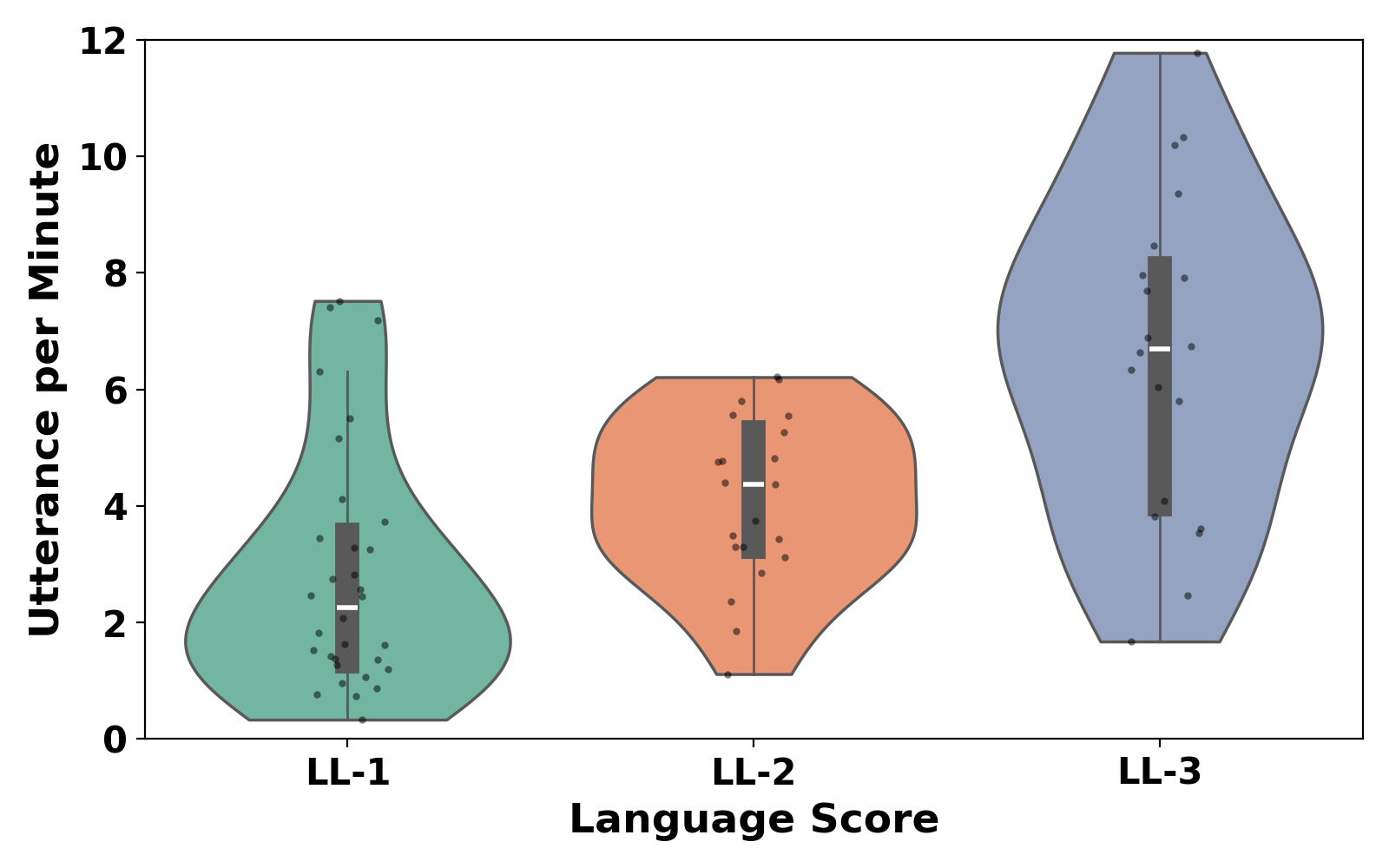}
    \vspace{-6.5mm}
    \caption{Utterance rate (utterances per minute) derived from the speaker diarization model by language level (LL-1 to LL-3) in the NLS study. The violin plot shows that the utterance rate increases with the language levels.}
    \vspace{-2.5mm}
    \label{fig:nls}
} \end{figure}

\paragraph{Observations} Figure~\ref{fig:nls} reveals a clear monotonic relationship between language level and utterance rate: the median number of utterances per minute increases steadily from LL-1 (pre-verbal communication) through LL-2 (first words) to LL-3 (word combinations). This trend indicates that children at more advanced spoken language levels are more vocally active, producing speech at a higher rate. Moreover, the variability in utterance rate is increased with language level, with LL-3 showing the most varied distribution. Overall, these results show that diarization-derived utterance rate captures a meaningful developmental signal consistent with expert-assigned language levels.

\subsection{Speech Production Development}

\paragraph{Application Setup} 
Our second application uses the PERCEPT-R corpus (Section~\ref{sec:benchmark}), which contains single-word rhotic-targeted speech samples from children with typical development and SSDs, each manually annotated for correct rhotic production. We apply our Whisper-Large and BabyHuBERT Rhotic/Derhotic classification model to investigate how the predicted probability of rhotic production varies with chronological age. Specifically, we restrict the analysis only to typically developing children, using recordings labeled ``PRE'' in the CRESULTSMOSAIC subset~\cite{ayala2023auditory}, and average the predictions from Whisper-Large and BabyHuBERT as the final output.

\begin{figure}[t] {
    \centering
    
    \includegraphics[width=\linewidth]{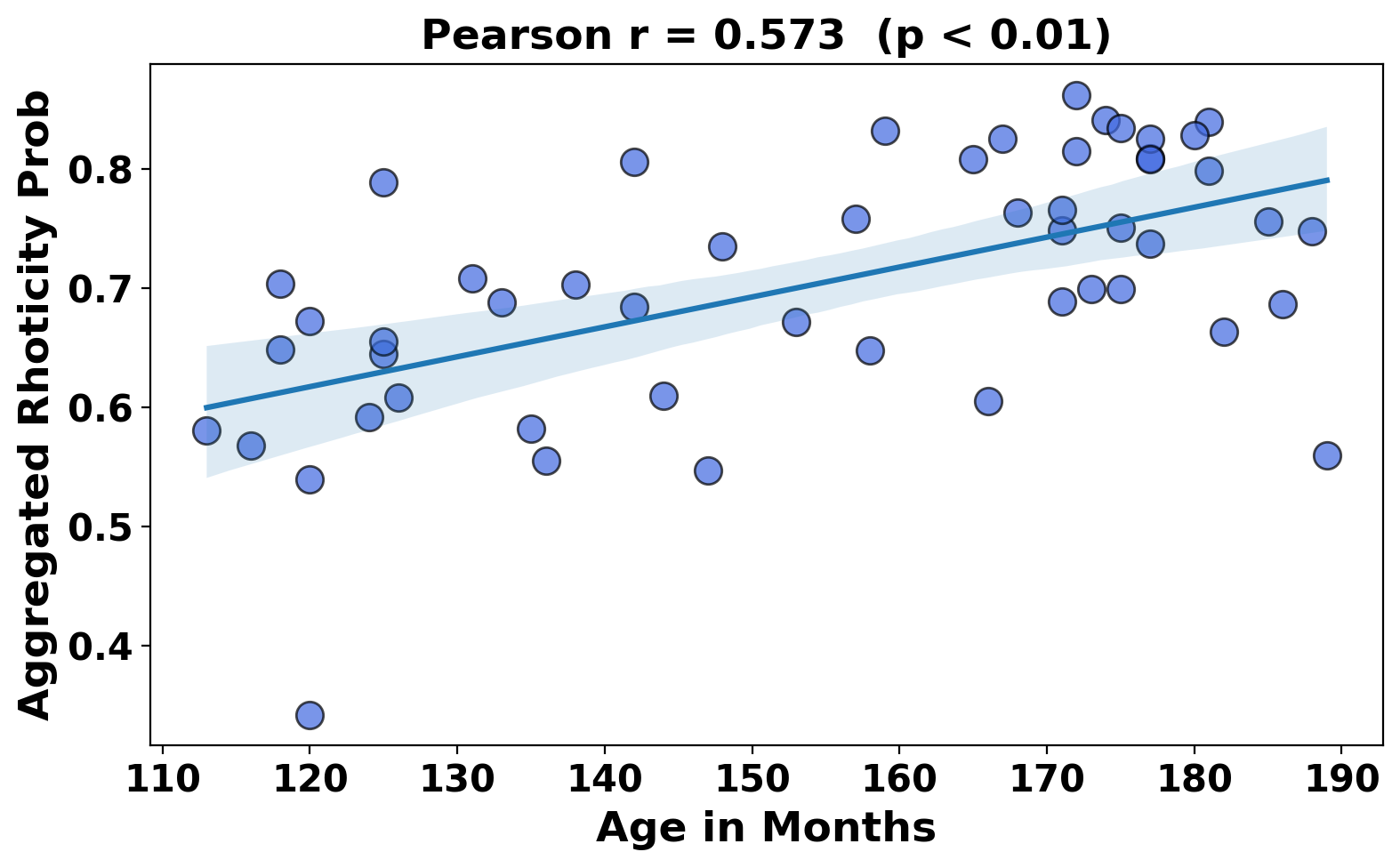}

    \vspace{-2mm}
    \caption{Correlation between model-predicted probability of correct rhotic production and chronic age in the CRESULTSMOSAIC subset of the PERCEPT-R dataset. Each dot represents a single participant, where the model-predicted probability is aggregated from all reading samples of a participant.}
    \vspace{-2mm}
    \label{fig:percet_r}
} \end{figure}

\paragraph{Observations} 
Figure~\ref{fig:percet_r} shows a moderate positive correlation between chronological age and the model-predicted probability of rhotic production in the CRESULTSMOSAIC subset of the PERCEPT-R ($r=0.573$). Specifically, younger children ($\sim$115-135 months) show larger variability, while predicted probabilities from older children centered more consistently around 0.75. Several lower-probability outliers remain present at older ages, highlighting continued inter-speaker variability. 
These results indicate the \texttt{ChildVox} provides models that can capture developmental changes in rhotic articulation across late childhood

\section{Conclusion and Future Work}

We introduce \texttt{ChildVox}, a benchmark characterizing the diverse acoustic signals through which children communicate from birth through school age, integrating more than 20 sub-tasks across 17 child-centered audio and speech datasets. 
We would highlight that the scope of the benchmark design is rarely presented in the literature.
Evaluated on a representative range of speech and audio foundation models, \texttt{ChildVox} supports downstream applications such as characterizing children's language levels and tracking speech production with age. In the future, we plan to expand it toward broader phoneme recognition~\cite{bharadwaj2026prism,guo2026huper}, additional recording environments~\cite{li2021analysis,feng2025egocentric,long2024babyview}, and to apply our benchmark models to enrich existing child audio corpora to support early screening~\cite{islam2024preliminary}, longitudinal monitoring~\cite{kalenkovich2025year}, and education~\cite{li2026k, wagner2025ohio} for children.

\section{Acknowledgement}

This work was supported by the Simons Foundation under Grant SFI-ARHUMAN-00004115-03 and Grant 655054. 

\section{Limitations}

While \texttt{ChildVox} facilitates the systematic evaluation of audio, speech, and LALMs on child-centered audios, several limitations remain.

\paragraph{Language, Demographics, and Task Coverage} The majority of datasets in \texttt{ChildVox} include most English-language recordings in speech categories, with limited representation of other languages or dialects. Specifically, our ASR evaluation is restricted to the English subset. Therefore, conclusions about model performance may not generalize to children speaking other languages, such as Mandarin~\cite{chua2023merlion} or Spanish~\cite{gildersleeve2008english}. Similarly, the demographic factors (e.g., educational background or developmental status) across the experimental datasets are, in many cases, not fully documented, which may introduce sampling biases that propagate into model evaluation and result interpretation. Moreover, we would consider integrating more spoken language understanding tasks as described in Dynamic-superb~\cite{huang2025dynamic,huang2024dynamic}.

\paragraph{Annotation variability} Many child-centered audio tasks in \texttt{ChildVox}, particularly affective vocalization classification (ReCANVo), cry-cause classification (Donate-a-Cry), and canonical syllables labeling (SpeechMaturity), involve inherently subjective categories with inter-rater disagreement. Therefore, reported classification scores may only reflect the ceiling imposed by annotation reliability.

\paragraph{Restricted set of foundation models} Although \texttt{ChildVox} includes several representative self-supervised, ASR-oriented, and LALMs, the development of audio foundation models is expanding rapidly. Specifically, we did not evaluate every recent open-source LALMs, such as GAMA~\cite{ghosh2024gama}, SALMONN~\cite{tang2024salmonn}, Step-Audio~\cite{huang2025step}, and Kimi-Audio~\cite{ding2025kimi}. Moreover, comparisons with proprietary baselines are limited to two Gemini Flash models in a zero-shot setting, and our zero-shot prompt approach did not cover different prompt-engineering techniques.

\section{Ethical Considerations}

Overall, all models released in this work are derived from publicly available sources, and \texttt{ChildVox} adheres to the scope of each dataset's license. Moreover, modeling children's speech can expose sensitive information about children, and to mitigate misuse, we plan to release code and most checkpoints under the Responsible AI License (RAIL). Users are expected to respect the privacy of data subjects and comply with applicable laws in their jurisdictions. We encourage the use of these models for research aimed at developing robust speech processing technology for and about children. Specifically, use of the models for clinical or diagnostic applications, surveillance, privacy-invasive applications, or commercial purposes is strictly prohibited. It is worth noting that models derived from data hosted using FluencyBank~\citep{ratner2018fluency}, part of the TalkBank system~\citep{macwhinney2007talkbank}, are subject to TalkBank's Ground Rules and distributed under CC BY-NC-SA 3.0.

\bibliography{refs}

@article{tager2009defining,
  title={Defining spoken language benchmarks and selecting measures of expressive language development for young children with autism spectrum disorders},
  author={Tager-Flusberg, Helen and Rogers, Sally and Cooper, Judith and Landa, Rebecca and Lord, Catherine and Paul, Rhea and Rice, Mabel and Stoel-Gammon, Carol and Wetherby, Amy and Yoder, Paul},
  journal={Journal of Speech, Language, and Hearing Research},
  volume={52},
  number={3},
  pages={643--652},
  year={2009}
}

@article{rocha2019open,
  title={An open access database for the evaluation of respiratory sound classification algorithms},
  author={Rocha, Bruno M and Filos, Dimitris and Mendes, Lu{\'\i}s and Serbes, Gorkem and Ulukaya, Sezer and Kahya, Yasemin P and Jakovljevic, Nik{\v{s}}a and Turukalo, Tatjana L and Vogiatzis, Ioannis M and Perantoni, Eleni and others},
  journal={Physiological measurement},
  volume={40},
  number={3},
  pages={035001},
  year={2019},
  publisher={IOP Publishing}
}

@inproceedings{koudounas2025voc2vec,
  title={voc2vec: A foundation model for non-verbal vocalization},
  author={Koudounas, Alkis and La Quatra, Moreno and Siniscalchi, Sabato Marco and Baralis, Elena},
  booktitle={ICASSP 2025-2025 IEEE International Conference on Acoustics, Speech and Signal Processing (ICASSP)},
  pages={1--5},
  year={2025},
  organization={IEEE}
}

@inproceedings{gong2022ssast,
  title={Ssast: Self-supervised audio spectrogram transformer},
  author={Gong, Yuan and Lai, Cheng-I and Chung, Yu-An and Glass, James},
  booktitle={Proceedings of the AAAI Conference on Artificial Intelligence},
  volume={36},
  number={10},
  pages={10699--10709},
  year={2022}
}

@article{lord2018autism,
  title={Autism spectrum disorder},
  author={Lord, Catherine and Elsabbagh, Mayada and Baird, Gillian and Veenstra-Vanderweele, Jeremy},
  journal={The lancet},
  volume={392},
  number={10146},
  pages={508--520},
  year={2018},
  publisher={Elsevier}
}

@inproceedings{ghosh2026audio,
 author = {Ghosh, Sreyan and Goel, Arushi and Kim, Jaehyeon and Kumar, Sonal and Kong, Zhifeng and Lee, Sang-gil and Yang, Chao-Han and Duraiswami, Ramani and Manocha, Dinesh and Valle, Rafael and Catanzaro, Bryan},
 booktitle = {Advances in Neural Information Processing Systems},
 doi = {10.52202/085713-1396},
 pages = {41819--41886},
 title = {Audio Flamingo 3: Advancing Audio Intelligence with Fully Open Large Audio Language Models},
 volume = {38, Main Conference},
 year = {2025}
}

@article{chu2024qwen2,
  title={Qwen2-audio technical report},
  author={Chu, Yunfei and Xu, Jin and Yang, Qian and Wei, Haojie and Wei, Xipin and Guo, Zhifang and Leng, Yichong and Lv, Yuanjun and He, Jinzheng and Lin, Junyang and others},
  journal={arXiv preprint arXiv:2407.10759},
  year={2024}
}

@article{ribeiro2021exploiting,
  title={Exploiting ultrasound tongue imaging for the automatic detection of speech articulation errors},
  author={Ribeiro, Manuel Sam and Cleland, Joanne and Eshky, Aciel and Richmond, Korin and Renals, Steve},
  journal={Speech Communication},
  volume={128},
  pages={24--34},
  year={2021},
  publisher={Elsevier}
}

@inproceedings{rao2024children,
  title={Children's Speech Emotion Recognition Using Feature Fusion-Based Deep Neural Networks},
  author={Rao, Dhulipalla Venkata and Radha, Kodali and Gudivaka, Sai Rekha and Polipogu, Pranuthi},
  booktitle={2024 IEEE International Conference on Intelligent Signal Processing and Effective Communication Technologies (INSPECT)},
  pages={1--6},
  year={2024},
  organization={IEEE}
}

@inproceedings{benway2022percept,
  title={PERCEPT-R: An Open-Access American English Child/Clinical Speech Corpus Specialized for the Audio Classification of /\:R/.},
  author={Benway, Nina and Preston, Jonathan L and Hitchcock, Elaine and Salekin, Asif and Sharma, Harshit and Byun, Tara McAllister},
  booktitle={INTERSPEECH},
  pages={3648--3652},
  year={2022}
}

@inproceedings{xu2023efficient,
  title={Efficient sequence transduction by jointly predicting tokens and durations},
  author={Xu, Hainan and Jia, Fei and Majumdar, Somshubra and Huang, He and Watanabe, Shinji and Ginsburg, Boris},
  booktitle={International Conference on Machine Learning},
  pages={38462--38484},
  year={2023},
  organization={PMLR}
}

@inproceedings{rekesh2023fast,
  title={Fast conformer with linearly scalable attention for efficient speech recognition},
  author={Rekesh, Dima and Koluguri, Nithin Rao and Kriman, Samuel and Majumdar, Somshubra and Noroozi, Vahid and Huang, He and Hrinchuk, Oleksii and Puvvada, Krishna and Kumar, Ankur and Balam, Jagadeesh and others},
  booktitle={2023 IEEE Automatic Speech Recognition and Understanding Workshop (ASRU)},
  pages={1--8},
  year={2023},
  organization={IEEE}
}

@inproceedings{xu2024exploring,
  title={Exploring Speech Foundation Models for Speaker Diarization in Child-Adult Dyadic Interactions},
  author={Xu, Anfeng and Huang, Kevin and Feng, Tiantian and Shen, Lue and Tager-Flusberg, Helen and Narayanan, Shrikanth},
  booktitle={Proc. Interspeech 2024},
  pages={5193--5197},
  year={2024}
}

@article{lavechin2026babar,
  title={BabAR: from phoneme recognition to developmental measures of young children's speech production},
  author={Lavechin, Marvin and Bergelson, Elika and Levy, Roger},
  journal={arXiv preprint arXiv:2603.05213},
  year={2026}
}

@article{lockhart2023infant,
  title={Infant cries convey both stable and dynamic information about age and identity},
  author={Lockhart-Bouron, Marguerite and Anikin, Andrey and Pisanski, Katarzyna and Corvin, Silo{\'e} and Cornec, Cl{\'e}ment and Papet, L{\'e}o and Levr{\'e}ro, Florence and Fauchon, Camille and Patural, Hugues and Reby, David and others},
  journal={Communications Psychology},
  volume={1},
  number={1},
  pages={26},
  year={2023},
  publisher={Nature Publishing Group UK London}
}

@inproceedings{pradhan2024my,
  title={My science tutor (myst)--a large corpus of children’s conversational speech},
  author={Pradhan, Sameer and Cole, Ronald and Ward, Wayne},
  booktitle={Proceedings of the 2024 Joint International Conference on Computational Linguistics, Language Resources and Evaluation (LREC-COLING 2024)},
  pages={12040--12045},
  year={2024}
}

@article{johnson2023recanvo,
  title={ReCANVo: A database of real-world communicative and affective nonverbal vocalizations},
  author={Johnson, Kristina T and Narain, Jaya and Quatieri, Thomas and Maes, Pattie and Picard, Rosalind W},
  journal={Scientific Data},
  volume={10},
  number={1},
  pages={523},
  year={2023},
  publisher={Nature Publishing Group UK London}
}

@article{cychosz2021vocal,
  title={Vocal development in a large-scale crosslinguistic corpus},
  author={Cychosz, Margaret and Cristia, Alejandrina and Bergelson, Elika and Casillas, Marisa and Baudet, Gladys and Warlaumont, Anne S and Scaff, Camila and Yankowitz, Lisa and Seidl, Amanda},
  journal={Developmental science},
  volume={24},
  number={5},
  pages={e13090},
  year={2021},
  publisher={Wiley Online Library}
}

@inproceedings{zhang25r_interspeech,
  title     = {{Employing self-supervised learning models for cross-linguistic child speech maturity classification}},
  author    = {Theo Zhang and Madurya Suresh and Anne Warluamont and Kasia Hitczenko and Alejandrina Cristia and Margaret Cychosz},
  year      = {2025},
  booktitle = {{Interspeech 2025}},
  pages     = {2825--2829}
}

@article{ding2025kimi,
  title={Kimi-audio technical report},
  author={Ding, Ding and Ju, Zeqian and Leng, Yichong and Liu, Songxiang and Liu, Tong and Shang, Zeyu and Shen, Kai and Song, Wei and Tan, Xu and Tang, Heyi and others},
  journal={arXiv preprint arXiv:2504.18425},
  year={2025}
}

@inproceedings{tang2024salmonn,
  title={Salmonn: Towards generic hearing abilities for large language models},
  author={Tang, Changli and Yu, Wenyi and Sun, Guangzhi and Chen, Xianzhao and Tan, Tian and Li, Wei and Lu, Lu and Ma, Zejun and Zhang, Chao},
  booktitle={International Conference on Learning Representations},
  year={2024}
}

@incollection{macwhinney2007talkbank,
  title={The talkbank project},
  author={MacWhinney, Brian},
  booktitle={Creating and digitizing language corpora: Volume 1: Synchronic databases},
  pages={163--180},
  publisher={Springer}
}

@article{huang2025step,
  title={Step-audio: Unified understanding and generation in intelligent speech interaction},
  author={Huang, Ailin and Wu, Boyong and Wang, Bruce and Yan, Chao and Hu, Chen and Feng, Chengli and Tian, Fei and Shen, Feiyu and Li, Jingbei and Chen, Mingrui and others},
  journal={arXiv preprint arXiv:2502.11946},
  year={2025}
}

@inproceedings{ghosh2024gama,
  title={Gama: A large audio-language model with advanced audio understanding and complex reasoning abilities},
  author={Ghosh, Sreyan and Kumar, Sonal and Seth, Ashish and Evuru, Chandra Kiran Reddy and Tyagi, Utkarsh and Sakshi, S and Nieto, Oriol and Duraiswami, Ramani and Manocha, Dinesh},
  booktitle={Proceedings of the 2024 Conference on Empirical Methods in Natural Language Processing},
  pages={6288--6313},
  year={2024}
}

@article{namasivayam2020speech,
  title={Speech sound disorders in children: An articulatory phonology perspective},
  author={Namasivayam, Aravind Kumar and Coleman, Deirdre and O’Dwyer, Aisling and Van Lieshout, Pascal},
  journal={Frontiers in psychology},
  volume={10},
  pages={2998},
  year={2020},
  publisher={Frontiers Media SA}
}

@article{lesaux2003development,
  title={The development of reading in children who speak English as a second language.},
  author={Lesaux, Nonie K and Siegel, Linda S},
  journal={Developmental psychology},
  volume={39},
  number={6},
  pages={1005},
  year={2003},
  publisher={American Psychological Association}
}

@article{li2026automated,
  title={Automated Analysis of Naturalistic Recordings in Early Childhood: Applications, challenges, and opportunities},
  author={Li, Jialu and Lavechin, Marvin and Fan, Xulin and McElwain, Nancy L and Cristia, Alejandrina and Garcia-Perera, Paola and Hasegawa-Johnson, Mark A},
  journal={IEEE Signal Processing Magazine},
  volume={42},
  number={6},
  pages={16--34},
  year={2026},
  publisher={IEEE}
}

@article{hitczenko2025speech,
  title={Speech Maturity Dataset: A cross-cultural corpus of naturalistic child and adult vocalizations},
  author={Hitczenko, Kasia and Peurey, Loann and Harvard, WN and Tey, Kai Jia and Seidl, Amanda and Semenzin, Chiara and Scaff, Camila and Lavechin, Marvin and Kelleher, Bridgette and Hamrick, Lisa and others},
  journal={OSF Preprints},
  year={2025}
}

@inproceedings{huang2024dynamic,
  title={Dynamic-superb: Towards a dynamic, collaborative, and comprehensive instruction-tuning benchmark for speech},
  author={Huang, Chien-yu and Lu, Ke-Han and Wang, Shih-Heng and Hsiao, Chi-Yuan and Kuan, Chun-Yi and Wu, Haibin and Arora, Siddhant and Chang, Kai-Wei and Shi, Jiatong and Peng, Yifan and others},
  booktitle={ICASSP 2024-2024 IEEE International Conference on Acoustics, Speech and Signal Processing (ICASSP)},
  pages={12136--12140},
  year={2024},
  organization={IEEE}
}

@inproceedings{huang2025dynamic,
  title={Dynamic-superb phase-2: A collaboratively expanding benchmark for measuring the capabilities of spoken language models with 180 tasks},
  author={Huang, Chien-yu and Chen, Wei-Chih and Yang, Shu-wen and Liu, Andy T and Li, Chen-An and Lin, Yu-Xiang and Tseng, Wei-Cheng and Diwan, Anuj and Shih, Yi-Jen and Shi, Jiatong and others},
  booktitle={International Conference on Learning Representations},
  year={2025}
}

@article{guo2026huper,
  title={HuPER: A Human-Inspired Framework for Phonetic Perception},
  author={Guo, Chenxu and Lian, Jiachen and Liu, Yisi and Huang, Baihe and Narayanan, Shriyaa and Cho, Cheol Jun and Anumanchipalli, Gopala},
  journal={arXiv preprint arXiv:2602.01634},
  year={2026}
}

@article{li2021analysis,
  title={Analysis of acoustic and voice quality features for the classification of infant and mother vocalizations},
  author={Li, Jialu and Hasegawa-Johnson, Mark and McElwain, Nancy L},
  journal={Speech communication},
  volume={133},
  pages={41--61},
  year={2021},
  publisher={Elsevier}
}

@article{islam2024preliminary,
  title={Preliminary technical validation of LittleBeats™: A multimodal sensing platform to capture cardiac physiology, motion, and vocalizations},
  author={Islam, Bashima and McElwain, Nancy L and Li, Jialu and Davila, Maria I and Hu, Yannan and Hu, Kexin and Bodway, Jordan M and Dhekne, Ashutosh and Roy Choudhury, Romit and Hasegawa-Johnson, Mark},
  journal={Sensors},
  volume={24},
  number={3},
  pages={901},
  year={2024},
  publisher={MDPI}
}

@article{kalenkovich2025year,
  title={A year of nouns from English-learning infants’ daily lives: The SEEDLingS-Nouns dataset: E. Kalenkovich et al.},
  author={Kalenkovich, Evgenii and Koorathota, Sharath and Tor, Shaelise and Amatuni, Andrei and Egan-Dailey, Shannon and Moore, Charlotte and Laing, Catherine and Garrison, Hallie and Baudet, Gladys and Bulgarelli, Federica and others},
  journal={Behavior Research Methods},
  volume={57},
  number={11},
  pages={298},
  year={2025},
  publisher={Springer}
}

@inproceedings{li2026k,
  title={K-function: Joint pronunciation transcription and feedback for evaluating kids language function},
  author={Li, Shuhe and Guo, Chenxu and Lian, Jiachen and Cho, Cheol Jun and Zhao, Wenshuo and Xu, Xiner and Jin, Ruiyu and Shi, Xiaoyu and Zhou, Xuanru and Zhou, Dingkun and others},
  booktitle={ICASSP 2026-2026 IEEE International Conference on Acoustics, Speech and Signal Processing (ICASSP)},
  pages={3231--3235},
  year={2026},
  organization={IEEE}
}

@article{bharadwaj2026prism,
  title={PRiSM: Benchmarking Phone Realization in Speech Models},
  author={Bharadwaj, Shikhar and Li, Chin-Jou and Kim, Yoonjae and Choi, Kwanghee and Yeo, Eunjung and Shim, Ryan Soh-Eun and Zhou, Hanyu and Boldt, Brendon and Jacome, Karen Rosero and Chang, Kalvin and others},
  journal={arXiv preprint arXiv:2601.14046},
  year={2026}
}

@book{lord2008autism,
  title={Autism diagnostic observation schedule (ADOS): Manual},
  author={Lord, Catherine and Rutter, Michael and DiLavore, Pamela C and Risi, Susan and Gotham, K and Bishop, SL and others},
  year={2008},
  publisher={Western Psychological Services Los Angeles}
}

@inproceedings{gemmeke2017audio,
  title={Audio set: An ontology and human-labeled dataset for audio events},
  author={Gemmeke, Jort F and Ellis, Daniel PW and Freedman, Dylan and Jansen, Aren and Lawrence, Wade and Moore, R Channing and Plakal, Manoj and Ritter, Marvin},
  booktitle={2017 IEEE international conference on acoustics, speech and signal processing (ICASSP)},
  pages={776--780},
  year={2017},
  organization={IEEE}
}

@inproceedings{feng2025egocentric,
  title={Egocentric Speaker Classification in Child-Adult Dyadic Interactions: From Sensing to Computational Modeling},
  author={Feng, Tiantian and Xu, Anfeng and Shi, Xuan and Bishop, Somer and Narayanan, Shrikanth},
  booktitle={Proc. Interspeech 2025},
  pages={2835-2839},
  year={2025}
}

@article{wagner2025ohio,
  title={The Ohio Child Speech Corpus},
  author={Wagner, Laura and Alghowinhem, Sharifa and Alwan, Abeer and Bowdrie, Kristina and Breazeal, Cynthia and Clopper, Cynthia G and Fosler-Lussier, Eric and Jamsek, Izabela A and Lander, Devan and Ramnath, Rajiv and others},
  journal={Speech Communication},
  volume={170},
  pages={103206},
  year={2025},
  publisher={Elsevier}
}

@inproceedings{eshky2018ultrasuite,
  title={UltraSuite: A Repository of Ultrasound and Acoustic Data from Child Speech Therapy Sessions},
  author={Eshky, Aciel and Ribeiro, Manuel Sam and Cleland, Joanne and Richmond, Korin and Roxburgh, Zoe and Scobbie, James M and Wrench, Alan},
  booktitle={Proc. Interspeech 2018},
  pages={1888--1892},
  year={2018}
}

@article{butler2022remote,
  title={Remote natural language sampling of parents and children with autism spectrum disorder: Role of activity and language level},
  author={Butler, Lindsay K and La Valle, Chelsea and Schwartz, Sophie and Palana, Joseph B and Liu, Cerelia and Peterman, Natalie and Shen, Lue and Tager-Flusberg, Helen},
  journal={Frontiers in Communication},
  volume={7},
  pages={820564},
  year={2022},
  publisher={Frontiers Media SA}
}

@article{oliveira2021circor,
  title={The CirCor DigiScope dataset: from murmur detection to murmur classification},
  author={Oliveira, Jorge and Renna, Francesco and Costa, Paulo Dias and Nogueira, Marcelo and Oliveira, Cristina and Ferreira, Carlos and Jorge, Al{\'\i}pio and Mattos, Sandra and Hatem, Thamine and Tavares, Thiago and others},
  journal={IEEE journal of biomedical and health informatics},
  volume={26},
  number={6},
  pages={2524--2535},
  year={2021},
  publisher={IEEE}
}

@inproceedings{zhang2021speechocean762,
  title={speechocean762: An Open-Source Non-Native English Speech Corpus for Pronunciation Assessment},
  author={Zhang, Junbo and Zhang, Zhiwen and Wang, Yongqing and Yan, Zhiyong and Song, Qiong and Huang, Yukai and Li, Ke and Povey, Daniel and Wang, Yujun},
  booktitle={Proc. Interspeech 2021},
  pages={3710--3714},
  year={2021}
}

@article{zhang2022sprsound,
  title={Sprsound: Open-source SJTU paediatric respiratory sound database},
  author={Zhang, Qing and Zhang, Jing and Yuan, Jiajun and Huang, Huajie and Zhang, Yuhang and Zhang, Baoqin and Lv, Gaomei and Lin, Shuzhu and Wang, Na and Liu, Xin and others},
  journal={IEEE Transactions on Biomedical Circuits and Systems},
  volume={16},
  number={5},
  pages={867--881},
  year={2022},
  publisher={IEEE}
}

@article{feng2025vox,
  title={Vox-Profile: A Speech Foundation Model Benchmark for Characterizing Diverse Speaker and Speech Traits},
  author={Feng, Tiantian and Lee, Jihwan and Xu, Anfeng and Lee, Yoonjeong and Lertpetchpun, Thanathai and Shi, Xuan and Wang, Helin and Thebaud, Thomas and Moro-Velazquez, Laureano and Byrd, Dani and others},
  journal={arXiv preprint arXiv:2505.14648},
  year={2025}
}

@article{ratner2018fluency,
  title={Fluency Bank: A new resource for fluency research and practice},
  author={Ratner, Nan Bernstein and MacWhinney, Brian},
  journal={Journal of fluency disorders},
  volume={56},
  pages={69--80},
  year={2018},
  publisher={Elsevier}
}

@article{ayala2023auditory,
  title={Auditory and somatosensory development for speech in later childhood},
  author={Ayala, Samantha A and Eads, Amanda and Kabakoff, Heather and Swartz, Michelle T and Shiller, Douglas M and Hill, Jennifer and Hitchcock, Elaine R and Preston, Jonathan L and McAllister, Tara},
  journal={Journal of Speech, Language, and Hearing Research},
  volume={66},
  number={4},
  pages={1252--1273},
  year={2023},
  publisher={American Speech-Language-Hearing Association}
}

@article{hu2022lora,
  title={Lora: Low-rank adaptation of large language models.},
  author={Hu, Edward J and Shen, Yelong and Wallis, Phillip and Allen-Zhu, Zeyuan and Li, Yuanzhi and Wang, Shean and Wang, Lu and Chen, Weizhu and others},
  journal={International Conference on Learning Representations},
  year={2022}
}

@inproceedings{pepino2021emotion,
  title={Emotion Recognition from Speech Using wav2vec 2.0 Embeddings},
  author={Pepino, Leonardo and Riera, Pablo and Ferrer, Luciana},
  booktitle={Proc. Interspeech 2021},
  pages={3400--3404},
  year={2021}
}

@article{chen2022wavlm,
  title={Wavlm: Large-scale self-supervised pre-training for full stack speech processing},
  author={Chen, Sanyuan and Wang, Chengyi and Chen, Zhengyang and Wu, Yu and Liu, Shujie and Chen, Zhuo and Li, Jinyu and Kanda, Naoyuki and Yoshioka, Takuya and Xiao, Xiong and others},
  journal={IEEE Journal of Selected Topics in Signal Processing},
  volume={16},
  number={6},
  pages={1505--1518},
  year={2022},
  publisher={IEEE}
}

@article{fan2024benchmarking,
  title={Benchmarking Children's ASR with Supervised and Self-supervised Speech Foundation Models},
  author={Fan, Ruchao and Shankar, Natarajan Balaji and Alwan, Abeer},
  journal={Interspeech},
  year={2024}
}

@article{ying2025benchmarking,
  title={Benchmarking training paradigms, dataset composition, and model scaling for child asr in espnet},
  author={Ying, Anyu and Shankar, Natarajan Balaji and Lin, Chyi-Jiunn and Shi, Mohan and Wang, Pu and Shim, Hye-jin and Arora, Siddhant and Alwan, Abeer and Watanabe, Shinji and others},
  journal={Workshop on Child Computer Interaction (WOCCI)},
  year={2025}
}

@article{singh2026causal,
  title={Causal analysis of ASR errors for children: Quantifying the impact of physiological, cognitive, and extrinsic factors},
  author={Singh, Vishwanath Pratap and Sahidullah, Md and Kinnunen, Tomi H},
  journal={Computer Speech \& Language},
  volume={95},
  pages={101859},
  year={2026},
  publisher={Elsevier}
}

@inproceedings{mcgonigle2024benchmarking,
  title={Benchmarking automatic speech recognition technology for natural language samples of children with and without developmental delays},
  author={McGonigle, Emma and VanDam, Mark and Wilkinson, Carol and Johnson, Kristina T},
  booktitle={2024 46th Annual International Conference of the IEEE Engineering in Medicine and Biology Society (EMBC)},
  pages={1--5},
  year={2024},
  organization={IEEE}
}

@inproceedings{li2024analysis,
  title={Analysis of self-supervised speech models on children’s speech and infant vocalizations},
  author={Li, Jialu and Hasegawa-Johnson, Mark and McElwain, Nancy L},
  booktitle={2024 IEEE International Conference on Acoustics, Speech, and Signal Processing Workshops (ICASSPW)},
  pages={550--554},
  year={2024},
  organization={IEEE}
}

@article{xu2023understanding,
  title={Understanding spoken language development of children with asd using pre-trained speech embeddings},
  author={Xu, Anfeng and Hebbar, Rajat and Lahiri, Rimita and Feng, Tiantian and Butler, Lindsay and Shen, Lue and Tager-Flusberg, Helen and Narayanan, Shrikanth},
  journal={Interspeech},
  year={2023}
}

@article{medin2025self,
  title={Self-Supervised Models for Phoneme Recognition: Applications in Children's Speech for Reading Learning},
  author={Medin, Lucas Block and Pellegrini, Thomas and Gelin, Lucile},
  journal={Interspeech},
  year={2024}
}

@article{sinha2026study,
  title={A study on the layer-wise transferability of self-supervised learning features for children’s speech processing tasks},
  author={Sinha, Abhijit and Kathania, Hemant Kumar and Kurimo, Mikko},
  journal={Speech Communication},
  pages={103392},
  year={2026},
  publisher={Elsevier}
}

@article{rose2014phonbank,
  title={The PhonBank project},
  author={Rose, Yvan and MacWhinney, Brian},
  journal={The Oxford Handbook of Corpus Phonology},
  pages={380--401},
  year={2014}
}

@article{long2024babyview,
  title={The BabyView camera: designing a new head-mounted camera to capture children’s early social and visual environments},
  author={Long, Bria and Goodin, Sarah and Kachergis, George and Marchman, Virginia A and Radwan, Samaher F and Sparks, Robert Z and Xiang, Violet and Zhuang, Chengxu and Hsu, Oliver and Newman, Brett and others},
  journal={Behavior Research Methods},
  volume={56},
  number={4},
  pages={3523--3534},
  year={2024},
  publisher={Springer}
}

@article{gildersleeve2008english,
  title={English speech sound development in preschool-aged children from bilingual English--Spanish environments},
  author={Gildersleeve-Neumann, Christina E and Kester, Ellen S and Davis, Barbara L and Pe{\~n}a, Elizabeth D},
  journal={Language, Speech, and Hearing Services in Schools},
  volume={39},
  number={3},
  pages={314--328},
  year={2008}
}

@inproceedings{chua2023merlion,
  title={MERLIon CCS Challenge: A English-Mandarin code-switching child-directed speech corpus for language identification and diarization},
  author={Chua, Victoria YH and Liu, Hexin and Garcia, Leibny Paola and Woon, Fei Ting and Wong, Jinyi and Zhang, Xiangyu and Khudanpur, Sanjeev and Khong, Andy WH and Dauwels, Justin and Styles, Suzy J},
  booktitle={Proc. Interspeech 2023},
  pages={4109--4113},
  year={2023}
}

@inproceedings{feng2024foundation,
  title={Foundation model assisted automatic speech emotion recognition: Transcribing, annotating, and augmenting},
  author={Feng, Tiantian and Narayanan, Shrikanth},
  booktitle={ICASSP 2024-2024 IEEE International Conference on Acoustics, Speech and Signal Processing (ICASSP)},
  pages={12116--12120},
  year={2024},
  organization={IEEE}
}

@article{xu2026end,
  title={End-to-End Joint ASR and Speaker Role Diarization with Child-Adult Interactions},
  author={Xu, Anfeng and Feng, Tiantian and Bishop, Somer and Lord, Catherine and Narayanan, Shrikanth},
  journal={IEEE Transactions on Audio, Speech and Language Processing},
  year={2026},
  publisher={IEEE}
}

@inproceedings{li2023towards,
  title={Towards robust family-infant audio analysis based on unsupervised pretraining of wav2vec 2.0 on large-scale unlabeled family audio},
  author={Li, Jialu and Hasegawa-Johnson, Mark and McElwain, Nancy L},
  booktitle={Interspeech},
  volume={2023},
  pages={1035},
  year={2023}
}

@article{charlot2025babyhubert,
  title={Babyhubert: Multilingual self-supervised learning for segmenting speakers in child-centered long-form recordings},
  author={Charlot, Th{\'e}o and Kunze, Tarek and Poli, Maxime and Cristia, Alejandrina and Dupoux, Emmanuel and Lavechin, Marvin},
  journal={arXiv preprint arXiv:2509.15001},
  year={2025}
}

\appendix

\section{Dataset Details}
\label{app:dataset}

\paragraph{CirCor}

The CirCor DigiScope Phonocardiogram Dataset~\cite{oliveira2021circor} is one publicly available pediatric heart-sound corpus, which includes 5,272 phonocardiogram recordings from 1,568 subjects aged 0–21 years.
Each subject is annotated by an expert cardiac physiologist with a murmur label drawn from present, absent, or unknown. We use the publicly released training/testing partition to classify these three labels. We segment each recording into non-overlapping 10-second windows. We further perform between-subject 5-fold cross-validation, partitioning the train data with 80/20 into train and development sets.

\paragraph{ICBHI}
ICBHI~\cite{rocha2019open} is a publicly available benchmark for respiratory sound analysis, targeting the recognition of crackles, wheezes, and respiratory conditions. It comprises 920 recordings collected from 126 subjects, accompanied by two complementary annotation schemes. The dataset provides cycle-level labels for 6,898 respiratory cycles, each categorized as normal, containing crackles, containing wheezes, or both. Similar to the CirCor dataset, we use the official ICBHI train/test split and apply 5-fold cross-validation over the training subjects to construct train/dev folds. For respiratory condition classification, recordings are segmented into 10-second windows with a 5-second hop, with diagnoses mapped to healthy, COPD, and other. For crackle and wheeze detection, we instead segment each recording according to the provided respiratory-cycle annotations. Segments shorter than 3 seconds (recording-level) or 0.5 seconds (cycle-level) are discarded.

\paragraph{SPRSound}
SPRSound~\cite{zhang2022sprsound} is a paediatric respiratory sound database, designed for automatic diagnosis. It contains 2683 records and 9089 respiratory sound events collected from 292 participants and annotated by 11 annotators. This dataset categorizes respiratory sound into two major labels: normal and adventitious. Specifically, there are continuous adventitious (CAS, such as Rhonchi, Wheeze, and Stridor) and discrete adventitious (DAS, such as Coarse Crackle and Fine Crackle) based on duration. We use the official SPRSound train/test split and apply 5-fold cross-validation over the training subjects to construct training and development folds. For respiratory sound classification, we use the record-level annotations provided with each recording, which are normal, CAS, DAS, CAS \& DAS, and poor quality.

\paragraph{Donate-a-cry} 
Donate-a-cry is a public infant cry audio corpus with labels for crying reasons. The audio samples in the corpus are collected with mobile devices and cleaned in the post-processing. There are 465 samples in the dataset, from children aged from 0 weeks to 2 years. The dataset is distributed with extreme imbalance crying reasons in belly\_pain (16), burping (8), discomfort (27), tired (24), and hungry (382). Therefore, we simplify the label as hungry and other.

\paragraph{What's in a cry?}
What's in a cry?~\cite{lockhart2023infant} is collected for a longitudinal infant crying study. 
The samples are recorded from 24 families with 10 girls and 14 boys. 
In our study, we focus on three primary cry causes, which are hunger, loneliness, and discomfort, discarding segments labeled as pain, other, or unknown due to limited sample availability. We merge consecutive cry segments within each recording when the inter-segment gap is less than 1 second to form continuous cry bouts, and retain only merged segments with a duration of at least 3 seconds. The resulting dataset is partitioned at the subject level using 5-fold cross-validation, where within each fold 80\% of the training subjects are used for training and the remaining 20\% for validation.

\paragraph{AudioSet}
AudioSet~\cite{gemmeke2017audio} is a large-scale ontology-driven collection of over one million YouTube video clips. In this benchmark, we focus on child sound event classification and filter data belonging to the following ten child-related audio categories: speaking, babbling, giggling, shouting, laughing, crying, whimpering, singing, playing, and music for children. We select the AudioSet-balanced dataset as the training set.

\paragraph{ReCANVo}
ReCANVo~\cite{johnson2023recanvo} is designed to investigate the informative expressions in nonverbal vocalizations from real-world interaction. The participants are 6 males and 2 females, aged from 6 to 23, with different developmental disorders. The speech samples from each speaker have at least 10 pieces with fewer than 10 words. Overall, the data has 7,077 audio samples, recorded in a home environment. In our benchmark, we use six expression categories: delighted, dysregulated, frustrated, request, selftalk, and social, while discarding samples that fall outside this label set. The dataset is partitioned at the utterance level using 5-fold cross-validation, where within each fold 80\% of the training and validation samples are used for training and the remaining 20\% for testing.

\paragraph{BabbleCor}
BabbleCor~\cite{cychosz2021vocal} is a crosslingual corpus of infants and children to investigate vocal development in typical speech. 52 participants are exposed to five different languages, including English, Spanish, Yêlí-Dnye, Tseltal Mayan, and bilingual Quechua-Spanish. Audio samples are clipped (400ms) from a day-long real-life recording for each child and annotated into five categories: canonical, non-canonical, crying, laughing, and junk. In our data processing, we use all five vocalization categories: crying, laughing, canonical, non-canonical, and junk, while discarding data marked as NO-LABEL. The dataset is partitioned at the subject level using 5-fold cross-validation, where within each fold 80\% of the training subjects are used for training and validation, and the remaining 20\% for testing. 

\paragraph{SpeechMaturity}
SpeechMaturity~\cite{hitczenko2025speech,zhang25r_interspeech} is also designed to investigate children's vocal development. Following BabbleCor, SpeechMaturity acquires clips from day-long children's speech recordings and gets clips annotated into the five categories. But SpeechMaturity is collected from a larger population size, containing 242,004 labeled vocalizations from more than 25 languages. Our data processing approach follows the same processing steps as stated in the BabbleCor dataset.

\paragraph{C-BESD}
C-BESD~\cite{rao2024children} is a bilingual dataset developed for children's emotion development research. The dataset contains 4200 utterances from 70 child speakers, aged from 6 to 12, and gender-balanced. Each speaker has 5 utterances for each of the 6 emotions (Happy, Neutral, Angry, Disgust, Fear, and Sad) in both Telugu and English. In our study, we focus on four primary emotion categories: angry, happy, neutral, and sad. This choice is similar to common emotion recognition benchmarks~\cite{feng2024foundation}. The dataset is partitioned at the speaker level using 5-fold cross-validation, where within each fold 80\% of the training subjects are used for training and validation, and the remaining 20\% for testing.

\paragraph{PERCEPT-R}
PERCEPT-R~\cite{benway2022percept} focuses on the speech of children from fully-rhotic American English with Rhotacism Speech Sound Disorder (RSSD).  PERCEPT-R contains 32.47 hours of citation speech recordings, comprising 105,232 word-level utterances. The 281 children (121 girls and 160 boys) are involved in speech recording, ranging from 72 months to 288 months. In data processing, we discard any utterances without a valid perceptual label. We restrict the data from the recordings including PRE-1 and DP-1 due to computational constraints. The dataset is partitioned at the subject level using 5-fold cross-validation, where within each fold 80\% of the training subjects are used for training and validation, and the remaining 20\% for testing.

\paragraph{SpeechOcean762}
SpeechOcean762~\cite{zhang2021speechocean762} is an open-source non-native English speech corpus for multi-granularity pronunciation assessment. It contains 5,000 utterances from 250 Mandarin-speaking learners (half children, half adults) with balanced gender and proficiency distributions. Each utterance is annotated by five experts to assess its accuracy, fluency, and prosody. Specifically, for accuracy and prosody, scores of 6 or below are mapped to ``Poor'', 7–8 to ``Nearly correct'', and 9–10 to ``Correct''. For fluency, scores below 5 are mapped to ``Intermittent'', 5–6 to ``Fluent in general'', and 7 or above to ``Fluent''. We adopt the official train and test splits, and further split the training set at the speaker level using 5-fold cross-validation.

\paragraph{UltraSuite}
UltraSuite~\cite{eshky2018ultrasuite} is a multimodal repository of synchronized ultrasound and acoustic data from child speech therapy sessions, containing 86 children (58 typically developing, 28 with SSDs) across three sub-datasets totaling 37.28 hours of audio and 14,456 utterances. Data spans six prompt types and includes longitudinal recordings across therapy stages for children with disorders such as phonological delay and childhood apraxia of speech.
In our benchmark, we focus on the typically developing children subset (core-uxtd) and restrict the data to Type A prompts, which consist of single-word utterances for word-level articulation analysis. We formulate a multi-label classification task targeting eight places of articulation: alveolar, bilabial, palatal, glottal, labiovelar, velar, dental, and labiodental, by mapping each target word to its constituent consonant phonemes via the CMU Pronouncing Dictionary.  We adopt the official train, dev, and test partitions provided with the dataset.

\paragraph{Natural Language Sampling (NLS)}
The dataset comprises 73 English-speaking child-parent dyadic interaction sessions that occurred in naturalistic or home settings. No child appears more than twice in the dataset. Most children are diagnosed with ASD and are minimally verbal. Recordings were made remotely by research personnel over Zoom. Domain experts assigned each child's spoken language levels to one of the three levels: pre-verbal (LL-1), first words (LL-2), or word combinations (LL-3). This definition was based on the prior study by ~\citet{tager2009defining}.

\paragraph{In-house ADOS2-Mod3} The ADOS~\cite{lord2008autism} is a semi-structured assessment protocol lasting 40–60 minutes, where ADOS-Module 3 (Mod3) targets verbally fluent children across 14 pre-defined activities. The in-house ADOS-Mod3 dataset~\cite{xu2026end} focuses on two interview-style components, Social Difficulties and Annoyance and Emotions. This produces 352 segments from 180 children (mean age 8.53 years, range 3–13). Roughly half received an ASD diagnosis, with most others diagnosed with ADHD or other neurodevelopmental or psychiatric conditions. 
We used the raw annotated data, whereas other works~\cite{xu2026end} applied additional preprocessing steps to clean the data.

\paragraph{MyST}
MyST~\cite{pradhan2024my} is one of the largest children's conversational speech corpora, containing 393 hours and 228,874 utterances from 1,371 third-to-fifth-grade students engaging in one-on-one spoken dialog with a virtual science tutor. The corpus includes a stratified train/dev/test split for the ASR experiment.

\paragraph{TinyVox}
TinyVox~\cite{lavechin2026babar}  is a large-scale cross-linguistic corpus of over 500,000 IPA-transcribed child vocalizations (387.7 hours) from 560 children aged 5–96 months, compiled and standardized from 31 PhonBank corpora spanning English, French, Portuguese, German, and Spanish, with a unified 57-phoneme target inventory mapped via phonological feature edit distance. The dataset provides a speaker-independent 80/10/10 train/validation/test split. In this study, we only perform the phoneme experiments with LALM, since the token spaces of the decoders for the evaluated ASR models are not directly compatible with phoneme recognition targets.

\section{Training Details}
\label{sec:training}

\subsection{Training Parameters}

For non-ASR tasks, during fine-tuning, SSAST models are trained with learning rates from ${1\times10^{-4}, 5\times10^{-4}}$, while other pre-trained speech models use learning rates from ${2\times10^{-4}, 5\times10^{-4}, 1\times10^{-3}}$. We use a learning rate of $1\times10^{-4}$ in fine-tuning Qwen2Audio and AudioFlamingo3. All experiments are trained for 10 epochs. We set the maximum audio duration of 10 seconds for most audio and speech classification tasks. We apply a batch size of 256 with a mini-batch size of 8 when fine-tuning LALMs. During training, we applied several augmentations to the input waveforms: Gaussian noise was added with a probability of 1.0, using an SNR range of 3–30 dB; time stretching was used with a probability of 1.0, with stretch rates ranging from 0.9 to 1.1; and polarity inversion was applied with a probability of 0.5.

For ASR fine-tuning, all models are trained with a learning rate of $1\times10^{-4}$ and a batch size of 32. We train for 5 epochs on the MyST dataset and 20 epochs on the ADOS dataset. Prior to fine-tuning, all transcripts are normalized using Whisper's \texttt{BasicTextNormalizer}, and the same normalization is applied to the model outputs during inference. Unlike the work in \cite{fan2024benchmarking}, we do not filter samples based on transcription length or the zero-shot Whisper-large-v3 WER, allowing for a more realistic and comprehensive evaluation of model performance.

\subsection{Training Architecture}

\paragraph{Encoder-based Fine-tuning} 
Prior works \cite{pepino2021emotion, feng2025vox} have shown that fine-tuning hidden representations from pre-trained encoder layers of speech foundation models can effectively adapt to a broad range of downstream speech tasks. Based on this, we adopt the architecture in Figure~\ref{fig:lora}, where a learnable weighted average over all encoder layers (both convolutional and transformer) is passed through a 1D point-wise convolution, followed by temporal averaging, and fed to a fully connected classifier. We further apply \texttt{LoRA} \cite{hu2022lora} to all fine-tuning experiments for parameter-efficient adaptation of the foundation model backbones, including for the ASR fine-tunings. However, due to implementation differences with wav2vec-LB and BabyHuBERT, we choose to fine-tune these two models only using hidden representations without LoRA.

\paragraph{LALMs} As stated in the main paragraph, we directly apply LoRA modules with rank 64 into the query, key, value, down-projection, and up-projection layers of the language model, and feedforward layers of the audio encoder. We use 4-bit quantization during training and normal 16-bit precision at inference.

\subsection{Training Hardware}

All experiments with public datasets are conducted on a High-Performance Computing (HPC) cluster. Most experiments are conducted with a single A40 or V100 GPU. Experiments with private data are conducted on a GPU server with 4 A6000 GPUs. Most experiments on encoder-based models require a fine-tuning time of less than 2 days, and experiments on LALMs require 4-5 days on an A40 GPU.

\section{Prompt Design}
\label{sec:prompt}
While LALMs are trained on generalized speech tasks, they may not have the domain knowledge about vocalizations/speech patterns specific to children.
Hence, in our SFT pipeline, we design system prompts for each dataset so as to include descriptions for each of the classes in the datasets. 
System prompt was first designed by the authors and then filtered using a frontier model (Gemini 3.1-pro).
System prompts corresponding to SFT of each dataset are as follows:

\subsection{SpeechMaturity}
\begin{PromptBlock}
Listen to the child audio recording and classify the vocalization type.
    Output only the label - no explanation or punctuation.

    Labels:
    - Crying: distress vocalizations, whimpers, or wails
    - Laughing: laughter or giggling sounds
    - Canonical: well-formed syllables with clear consonant-vowel structure (e.g., ``ba'', ``da'')
    - Non-Canonical: vocalizations lacking clear consonant-vowel structure (e.g., squeals, growls, vowel-only sounds)
    - Junk: non-speech noise, silence, or unintelligible audio
\end{PromptBlock}
\subsection{BabbleCor}
\begin{PromptBlock}
    Listen to the child audio recording and classify the vocalization type.
    Output only the label - no explanation or punctuation.

    Labels:
    - Crying: distress vocalizations, whimpers, or wails
    - Laughing: laughter or giggling sounds
    - Canonical: well-formed syllables with clear consonant-vowel structure (e.g., ``ba'', ``da'')
    - Non-Canonical: vocalizations lacking clear consonant-vowel structure (e.g., squeals, growls, vowel-only sounds)
    - Junk: non-speech noise, silence, or unintelligible audio
\end{PromptBlock}
\subsection{PERCEPT-R}
\begin{PromptBlock}
    Listen to the child reading a word. Evaluate the pronunciation of the 'r' sound specifically.
    Output only the label - no explanation or punctuation.

    Labels:
    - Rhotic: the 'r' sound is produced correctly with appropriate rhoticity
    - Derhotic: the 'r' sound is missing, distorted, or replaced (e.g., substituted with 'w')
\end{PromptBlock}
\subsection{SpeechOcean762-Prosody}
\begin{PromptBlock}
    Listen to the child reading a sentence. Evaluate the prosody - focus on intonation, stress, and rhythm.
    Output only the label - no explanation or punctuation.

    Labels:
    - Poor intonation: flat, monotone, or clearly unnatural intonation patterns
    - Nearly correct intonation: mostly natural prosody with minor errors
    - Correct intonation: natural, native-like intonation and rhythm throughout
\end{PromptBlock}
\subsection{SpeechOcean762-Fluency}
\begin{PromptBlock}
    Listen to the child reading a sentence. Evaluate the fluency - focus on flow, pausing, and hesitations.
    Output only the label - no explanation or punctuation.

    Labels:
    - Intermittent or little fluent: frequent pauses, hesitations, or disruptions in speech flow
    - Fluent in general: mostly smooth delivery with occasional minor disfluencies
    - Fluent: smooth, continuous speech with no notable disfluencies
\end{PromptBlock}
\subsection{SpeechOcean762-Accuracy}
\begin{PromptBlock}
    Listen to the child reading a sentence. Evaluate the pronunciation accuracy - focus on correct phoneme production.
    Output only the label - no explanation or punctuation.

    Labels:
    - Poor or Understandable: many mispronunciations; difficult to understand
    - Good: mostly accurate pronunciation with some errors
    - Excellent: highly accurate, native-like pronunciation throughout
\end{PromptBlock}
\subsection{CirCor}
\begin{PromptBlock}
    Listen to the heart sound recording. Determine whether a cardiac murmur is present.
    Output only the label - no explanation or punctuation.

    Labels:
    - Absent: no murmur detected; normal heart sounds
    - Present: a murmur is clearly audible
    - Unknown: audio quality or ambiguity prevents a confident determination
\end{PromptBlock}
\subsection{ICBHI-Crackles}
\begin{PromptBlock}
    Listen to the lung sound recording. Determine whether crackles are present.
    Output only the label - no explanation or punctuation.

    Labels:
    - Crackles: discontinuous, explosive sounds (fine or coarse crackles) are audible
    - No crackles: no crackle sounds detected
\end{PromptBlock}
\subsection{ICBHI-Wheezes}
\begin{PromptBlock}
    Listen to the lung sound recording. Determine whether wheezes are present.
    Output only the label - no explanation or punctuation.

    Labels:
    - Wheezes: continuous, high-pitched whistling sounds are audible
    - No wheezes: no wheeze sounds detected
\end{PromptBlock}
\subsection{SPRSound}
\begin{PromptBlock}
    Listen to the lung sound recording. Classify the respiratory sound pattern.
    Output only the label - no explanation or punctuation.

    Labels:
    - Normal: clear breath sounds with no adventitious sounds
    - CAS: continuous adventitious sounds only (e.g., wheezes, rhonchi)
    - DAS: discontinuous adventitious sounds only (e.g., crackles)
    - CAS & DAS: both continuous and discontinuous adventitious sounds present
    - Poor Quality: recording is too noisy or unclear for reliable classification
\end{PromptBlock}
\subsection{ReCANVo}
\begin{PromptBlock}
    Listen to the child audio recording and classify the communicative intent or emotional state of the vocalization.
    Output only the label - no explanation or punctuation.

    Labels:
    - Delighted: joyful, excited, or positively expressive sounds
    - Dysregulated: distressed, overwhelmed, or emotionally dysregulated vocalizations
    - Frustrated: sounds expressing frustration or dissatisfaction
    - Request: vocalizations used to request an object, action, or attention
    - Selftalk: vocalizations directed inward, not toward a social partner
    - Social: vocalizations directed toward engaging or interacting with others
\end{PromptBlock}
\subsection{What's in a cry?}
\begin{PromptBlock}
    Listen to the infant cry recording and identify the most likely cause of the cry.
    Output only the label - no explanation or punctuation.

    Labels:
    - Hunger: cry pattern associated with feeding need
    - Loneliness: cry pattern associated with desire for social contact or comfort. Loneliness was defined as the infant seeking contact or attention
    - Discomfort: cry pattern associated with physical discomfort. Examples of discomfort included cold, fever, and full diaper.
\end{PromptBlock}
\subsection{MyST}
\begin{PromptBlock}
    Listen to the child audio recording and transcribe exactly what is spoken.
    Output only the transcript - no labels, explanations, punctuation beyond what is spoken, or formatting.
    If the audio is unintelligible, output: [unintelligible]
\end{PromptBlock}
\subsection{TinyVox}
\begin{PromptBlock}
    Listen to the child speaking a single word. Transcribe the pronunciation using standard phoneme symbols (IPA).
    Output only the phoneme sequence - no labels, explanations, or formatting.
    Example format (IPA): /b ae t/
\end{PromptBlock}

\section{Additional Whisper Results}
\label{sec:additional_benchmark}

Additional results on Whisper-base and Whisper-small are presented in Table~\ref{tab:benchmark_results_whisper}.

\begin{table*}[ht]
    \centering
    \caption{Additional benchmark performance across physiological sounds, vocalization, canonical syllables, and speech from children on each complete dataset. 
    Whisper-B and Whisper-S are Whisper-Base and Whisper-Small, respectively.
    We report the average Macro-F1 score (higher the better).}
    \small
    \setlength{\tabcolsep}{6pt}
    \begin{tabular}{c l *{2}{S[table-format=1.3]}}
    \toprule
    {Task Category} & {Dataset} & {Whisper-B} & {Whisper-S} \\
    \midrule
     & {CirCor} & 0.499 & 0.546 \\
     & {ICBHI-crackles} & 0.581 & 0.597 \\
     Physiological & {ICBHI-wheezes} &  0.614 & {0.621} \\
     Sounds & {ICBHI-Condition} & 0.473 & {0.483} \\
     & {SPRSound} & 0.345 & 0.362 \\
    
    \midrule
    \multirow{3}{*}{Vocalization}
     & {AudioSet-Child} & 0.598 & {0.634} \\
     & {What's in a cry?} & {0.333} & {0.326} \\
     & {donate-a-cry} & 0.467 & {0.528} \\
    \midrule
    \multirow{3}{*}{\shortstack{Canonical\\Syllables}}
     & {ReCANVo} & 0.414 & 0.415 \\
     & {Speech Maturity} & 0.625 & 0.634 \\
     & {BabbleCor} & 0.546 & {0.596} \\
    \midrule
    \multirow{7}{*}{Speech}
     & {NLS-Speech Intelligibility} & 0.635 & {0.658} \\
     & {Percept-R} & 0.794 & 0.807 \\
     & {Ultrasuite} & {0.923} & {0.926} \\
     & {SpeechOcean-Fluency} & 0.625 & {0.626} \\
     & {SpeechOcean-Accuracy} & 0.605 & {0.628} \\
     & {SpeechOcean-Prosody} & 0.701 & 0.704 \\
     & {C-BESD} & 0.596 & 0.593 \\
    \bottomrule
    \end{tabular}
    \label{tab:benchmark_results_whisper}
    \vspace{-2mm}
\end{table*}

\section{Proprietary Baselines}
\label{sec:proprietary}
\begin{figure}[t] {
    \centering
    
    \includegraphics[width=\linewidth]{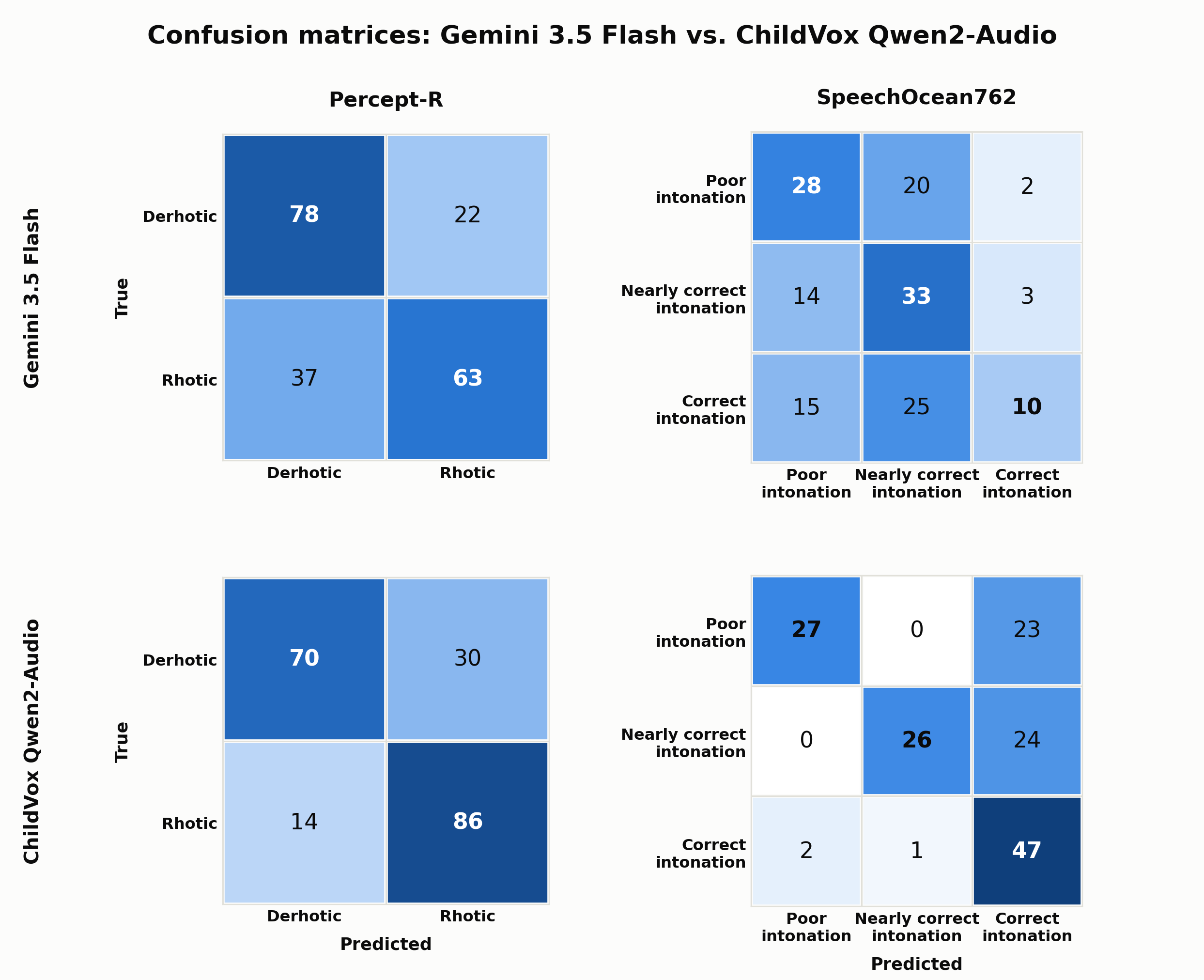}

    \vspace{-2mm}
    \caption{Confusion matrices of Gemini 3.5 Flash vs. ChildVox Qwen2-Audio on PERCEPT-R and SpeechOcean Intonation predictions.}
    \vspace{-2mm}
    \label{fig:confusion}
} \end{figure}

We apply both Gemini 2.5 Flash and Gemini 3.5 Flash APIs on 5 selected public datasets in the main paragraph. We apply a temperature of 0.1, 0.5, and 1.0 in prompting both models. We run the prompt experiment on these two proprietary baselines for a single run due to the budget limitation. Figure~\ref{fig:confusion} shows the confusion matrices of Gemini 3.5 Flash vs. ChildVox Qwen2-Audio on PERCEPT-R and SpeechOcean Intonation predictions.
Specifically, on PERCEPT-R, Gemini 3.5 Flash over-predicts derhotic, while our ChildVox Qwen-Audio 2 has better rhotic prediction. On SO-Prosody, Gemini fails toward the middle class, and ChildVox Qwen2-Audio instead over-predicts Correct intonation.

\section{Generative AI Use Disclaimer}
Generative AI tools were used solely for grammar checking and polishing the writing of the manuscript. All research concepts, experimental design, implementations, analyses, and reported results were developed and conducted by the authors. The conceptual framing, methodology, and scientific contributions of this work are entirely human-generated.

\end{document}